\documentclass[nofootinbib, showkeys, unsortedaddress, superscriptaddress, notitlepage]{revtex4-1}
\usepackage{graphicx}
\usepackage{amsmath}
\usepackage{amsthm}
\usepackage{dsfont}
\usepackage{mathptmx}
\usepackage{geometry}
\usepackage{booktabs}
\usepackage{array}
\newcolumntype{L}[1]{>{\raggedright\let\newline\\\arraybackslash\hspace{0pt}}m{#1}}
\newcolumntype{C}[1]{>{\centering\let\newline\\\arraybackslash\hspace{0pt}}m{#1}}
\newcolumntype{R}[1]{>{\raggedleft\let\newline\\\arraybackslash\hspace{0pt}}m{#1}}

\newcommand{\RR}{\mathds{R}}

\newcommand{\ZZ}{\mathds{Z}}
\newcommand{\NN}{\mathds{N}}
\newcommand{\dd}{\mathrm{d}}
\newcommand{\TT}{\mathds{T}}
\newcommand{\PP}{\mathds{P}}
\newcommand{\HH}{\mathds{H}}
\newcommand{\eps}{\varepsilon}
\newcommand{\uu}{u_{\text{I}}}
\newcommand{\pp}{p_{\text{I}}}
\newcommand{\norm}[1]{\left\lVert#1\right\rVert}

\newcommand{\lws}{
\vspace{1mm}
\hrule
\vspace{1mm}}

\usepackage{newfloat,algcompatible}
\usepackage{etoolbox}

\AtEndEnvironment{algorithm}{\noindent\hrulefill\par\nobreak\vskip-8pt}

\DeclareFloatingEnvironment[
    fileext=loa,
    listname=List of Algorithms,
    name=ALGORITHM,
    placement=tbhp,
]{algorithm}

\AtBeginDocument{
\heavyrulewidth=.08em
\lightrulewidth=.05em
\cmidrulewidth=.03em
\belowrulesep=.65ex
\belowbottomsep=0pt
\aboverulesep=.4ex
\abovetopsep=0pt
\cmidrulesep=\doublerulesep
\cmidrulekern=.5em
\defaultaddspace=.5em
}

\makeatletter
\def\p@subsection{}
\makeatother

\usepackage{hyperref}

\begin{document}

\title{Spontaneous Symmetry Breaking for Extreme Vorticity and Strain in the 3D Navier-Stokes Equations}

\author{Timo Schorlepp}
\email{Timo.Schorlepp@rub.de}
\affiliation{Institute for Theoretical Physics I, Ruhr-University Bochum,
               Universit\"atsstrasse 150, 44801 Bochum, Germany}

\author{Tobias Grafke}
\affiliation{Mathematics Institute, University of Warwick, Coventry CV4 7AL, United Kingdom}

\author{Sandra May}
\affiliation{Department of Mathematics, TU Dortmund University, Vogelpothsweg 87, 44227 Dortmund, Germany}

\author{Rainer Grauer}
\affiliation{Institute for Theoretical Physics I, Ruhr-University Bochum,
               Universit\"atsstrasse 150, 44801 Bochum, Germany}

\date{\today}

\begin{abstract}
  We investigate the spatio-temporal structure of the most likely
  configurations realising extremely high vorticity or strain in the
  stochastically forced 3D incompressible Navier-Stokes equations. Most
  likely configurations are computed by numerically finding the
  highest probability velocity field realising an extreme constraint
  as solution of a large optimisation problem. High-vorticity
  configurations are identified as pinched vortex filaments with
  swirl, while high-strain configurations correspond to counter-rotating
  vortex rings. We additionally observe that the most likely
  configurations for vorticity and strain spontaneously break their
  rotational symmetry for extremely high observable values. Instanton calculus
  and large deviation theory allow us to show that these maximum
  likelihood realisations determine the tail probabilities of the
  observed quantities. In particular, we are able to demonstrate that
  artificially enforcing rotational symmetry for large strain
  configurations leads to a severe underestimate of their probability,
  as it is dominated in likelihood by an exponentially more likely
  symmetry broken vortex-sheet configuration.
\end{abstract}

\maketitle

\tableofcontents


\section{Introduction and Motivation}

Turbulence is characterised by its tendency to intermittently
dissipate energy in very localised and intense events. These extreme
events dominate the statistics of quantities such as high order
structure functions, and are ultimately responsible for the anomalous
scaling of fully developed turbulent flows. It is generally believed
that short bursts of intense vortex stretching are the mechanism for
the formation of these events.

Taking this as starting point, in this paper we address the question:
What structures are naturally generated in the 3D incompressible
Navier-Stokes equations (NSE) to realise events of extreme vortex
stretching, strain production
and energy dissipation? For this, we are concentrating on small-scale
structures that lead to extreme values of the fluid vorticity or its
strain. Concretely, we set out to compute the most likely
configuration (for a given large-scale stochastic forcing) that
realises a large vorticity or strain value at a single point within
the domain, at an instantaneous moment in time, and how the velocity
field configuration around this point facilitates the extreme burst.

This question has been discussed in the literature, starting with
Novikov~\cite{novikov:1993, mui-dommermuth-novikov:1996}, and more
recent works that explored extreme vorticity and strain events in very
large turbulent simulations~\cite{yeung-zhai-sreenivasan:2015,
  buaria-pumir-bodenschatz-etal:2019}. These attempts, which solely
rely on brute-force direct numerical simulations (DNS), have the
intrinsic complication that any extreme realisation of an observable
will necessarily be very rare, and thus hard to observe. Therefore,
exploring extreme events not only requires high numerical resolution,
but further extremely large data-sets, most of which are wasted
because they do not exhibit the desired event. On this basis, we
instead employ specific rare event techniques~\cite{sapsis:2021}, in
particular stochastic field theory and instanton
calculus~\cite{grafke-grauer-schaefer:2015}, or equivalently, sample
path large deviation theory~\cite{freidlin-wentzell:2012}. The two are
intimately connected~\cite{grafke-grauer-schaefer:2015,
  grafke-vanden-eijnden:2019}, and have proven successful in related
fields, such as extreme shocks in Burgers
turbulence~\cite{balkovsky-falkovich-kolokolov-etal:1997,
  chernykh-stepanov:2001, grafke-grauer-schaefer-etal:2015}, extreme
surface heights in the Kardar-Parisi-Zhang~(KPZ)
equation~\cite{meerson-katzav-vilenkin:2016}, in ocean waves and
tsunamis~\cite{dematteis-grafke-onorato-etal:2019,
  tong-vanden-eijnden-stadler:2020}, or extreme mechanical forces in
grid-generated turbulence~\cite{lestang-bouchet-leveque:2020}.  The
key idea is to replace the inefficient naive sampling approach by a
deterministic optimisation problem that yields the maximum likelihood
trajectory of the system that leads to a prescribed rare outcome. The
advantage of this method is the fact that it yields the best estimate
of the \emph{typical} extreme event in the limit of it becoming
increasingly rare, which is the limit we are most interested in, and
at the same time also the regime that is hardest to reach via DNS.

As we will discuss later, instanton techniques not
only allow for the computation of the limiting most likely path to
obtain an extreme event, but further yield estimates for the
exponential tail scaling of the observable's probability density
function (PDF). A concrete prediction of our results is the fact that
intuitive rotationally symmetric realisations of extreme vorticity
outcomes (namely, vortex tubes/filaments) or extreme strain outcomes
(namely, colliding or contracting vortex rings) are not necessarily the
most likely way to reach extreme values, even if the observable
exhibits rotational symmetry. In fact we present that the rotationally
symmetric events become subdominant, particularly for large positive
strain values, and are dominated in probability by asymmetric field
configurations. In other words, the stochastic instanton undergoes
spontaneous symmetry breaking, and the corresponding action exhibits a
dynamical phase transition similar to what is observed e.g. in the KPZ
equation~\cite{janas-kamenev-meerson:2016}.

This paper is organised as follows: We discuss the instanton
approach, as applied to the NSE, in
section~\ref{sec:instantons-3d-navier} by first introducing the
instanton formalism in
section~\ref{sec:instantons-3d-navier}~\ref{sec:acti-minim-inst} in general and
subsequently applying it to the NSE in
section~\ref{sec:instantons-3d-navier}~\ref{sec:instanton-eq-nse-obsv}, where
we also explain our conditioning on vorticity and strain. The
numerical implementation of the corresponding optimisation problem is discussed in
section~\ref{sec:instantons-3d-navier}~\ref{sec:numerical-procedure}. In
section~\ref{sec:results}~\ref{sec:extr-vort-events} we show the most
likely configuration for extreme vorticity events as obtained by the
numerical solution of the instanton problem. Section~\ref{sec:results}~\ref{sec:extr-stra-events} presents analogous results
for extreme strain events. We will discuss
the implication of these results on
the likelihood and PDF tail scaling in
section~\ref{sec:results}~\ref{sec:extr-event-prob} and then conclude
with section~\ref{sec:conclusion}.
The supplemental material of this paper includes additional,
detailed information on the numerical optimisation methods that have been
used to generate the results of this paper.

\section{Instantons for the 3D Navier-Stokes equations}
\label{sec:instantons-3d-navier}

The 3D incompressible NSE on a domain
$\Omega\subset\RR^3$, given by
\begin{equation}
  \begin{cases}
    \partial_t u + (u\cdot\nabla)u = -\nabla P + \nu \Delta u + \eta\,,\\
    \nabla\cdot u = 0\,,\\
    u(\cdot, -T) = u_0\,,
  \end{cases}
  \label{eq:nse-initial}
\end{equation}
describe the spatio-temporal evolution of a velocity field $u:
\Omega\times [-T,0] \mapsto \RR^3$, where $P(x,t)$ is the pressure
field, $\eta(x,t)$ is the stochastic forcing term, $\nu>0$ is the
kinematic viscosity and~$u_0$ is a deterministic initial condition.
We restrict ourselves to a periodic domain $\Omega = [0, l]^3$,
and consider a white-in-time, spatially stationary and solenoidal
Gaussian forcing acting only on large scales as specified
by the spatial covariance $\chi:\Omega \to \RR^{3 \times 3}$:
\begin{equation}
	\left< \eta(x,t) \eta^\top(x',t') \right> = \chi(x-x') \delta(t - t')\,.
	\label{eq:forcing-covariance-general}
\end{equation}
Additionally requiring the forcing to be statistically isotropic
reduces the possible forms of $\chi$ to~\cite{robertson:1940}
\begin{equation}
\chi(x) = f\left( \lVert x\rVert \right) \mathrm{Id} +
\frac{1}{2} \lVert x\rVert f'\left( \lVert x\rVert \right)
\left[\mathrm{Id} - \frac{x x^\top}{\lVert x\rVert^2} \right]\,,
\end{equation}
where $\mathrm{Id}\in\RR^{3\times3}$ denotes the identity matrix on
$\RR^3$, and $f: [0, \infty) \to \RR$ is an arbitrary function, which
we choose as
\begin{align}
f(r) = \chi_0 \exp \left\{-\frac{r^2}{2\lambda^2}\right\}
\end{align}
for simplicity, with a correlation length~$\lambda$ of the order
of the domain size~$l$.

Extreme events in the NSE have been explored extensively in the literature.
Particularly worth mentioning in connection with the instanton calculus
is the work of Novikov~et~al.~\cite{novikov:1993,
  mui-dommermuth-novikov:1996}. They considered the conditionally
averaged vorticity field, i.e.~the average realisation of the
vorticity field conditioned on a specific outcome of vorticity
$\omega(x,t=0)$ at a given point $x$. These fields, parametrised by
$\omega$, were obtained by performing many DNS, and averaging
conditioned on the intended outcome. This procedure is closely
related to the filtering approach~\cite{grafke-grauer-schaefer:2013}
discussed in section~\ref{sec:results}~\ref{sec:extr-vort-events} and
demonstrates the relevance of instanton solutions in real flows.

The structure of instanton solutions is of particular importance.
As an example serves the observation that the rotational symmetric vorticity
instanton in the two-dimensional NSE has no relevance
at all \cite{falkovich-lebedev:2011}. Only taking into account symmetry
breaking angle dependent contributions results in an effective action
suitable for the instanton calculus.

In the case of the KPZ equation, symmetry breaking (or dynamical phase
transition) has been demonstrated as the mechanism to generate the relevant
instanton for obtaining the correct tail asymptotics
\cite{janas-kamenev-meerson:2016}. Here, we make similar observations:
symmetry breaking is essential to compute the relevant instanton
with a pancake or sheet like structure
(see figures~\ref{fig:vorticity-result} and \ref{fig:strain-result}).
Whether these structures are related to the recently discovered confined vortex
surfaces~\cite{migdal:2021} and the tangential discontinuity of vortex
sheets~\cite{agafontsev-kuznetsov-mailybaev:2021}
poses a challenging question.

\subsection{Stochastic action and minimisers}
\label{sec:acti-minim-inst}

In this section, we briefly and formally introduce the instanton
formalism for stochastic partial differential equations (SPDE) and
comment on the applicability of the method in the context of Navier-Stokes
turbulence to compute maximum likelihood space-time realisations of extreme events.
For a generic SPDE for $u:\Omega \times [-T,0] \to \RR^3$
\begin{align}
\begin{cases}
\label{eq:generic-SPDE}  
  \partial_t{u}(x,t) + N(u(\cdot, t))(x) = \sqrt{\eps} \eta(x,t)\,, \\
u(\cdot,  -T) = u_0\,,
\end{cases}
\end{align}
with a Gaussian forcing correlated according to~(\ref{eq:forcing-covariance-general})
and noise strength $\eps > 0$, expectations of a functional $F$ with respect to the
process $u$ can formally be computed as a path integral
\begin{align}
\left< F[u] \right> = \int D \eta \; F[u[\eta]] e^{-\frac{1}{2\eps} \int_{-T}^0
\left(\eta, \chi^{-1} * \eta \right)_{L^2(\Omega, \RR^3)} \mathrm{d} t} =
\int_{u(\cdot, -T) = u_0} Du \; F[u] J[u] e^{- \frac{1}{\eps} S[u]}\,,
\label{eq:path-integral}
\end{align}
where $*$ denotes spatial convolution and $\chi^{-1}$ is the
convolutional inverse of the forcing correlation function $\chi$.
The Jacobian $J[u]$ is given by
$J[u] = \exp \big\{ \frac{1}{2} \int_{-T}^0 \mathrm{tr\;} \nabla N(u)
\;\mathrm{d}t \big\}$ and $S$ is the classical
Onsager-Machlup~\cite{machlup-onsager:1953} or
Freidlin-Wentzell~\cite{freidlin-wentzell:2012} action functional
\begin{align}
S[u] = \int_{-T}^0 {\cal L}(u, \partial_t{u}) \mathrm{d}t = \frac{1}{2}
\int_{-T}^0 \left(\partial_t{u} + N(u), \chi^{-1} * [\partial_t{u} + N(u)] \right)_{L^2(\Omega, \RR^3)} \dd t
\label{eq:om-action}
\end{align}
of the process~$u$. In the case of a degenerate forcing, as in our
specific application, we set $S[u] = + \infty$ if the trajectory does
not lie in the image of the spatial convolution with~$\chi$.
Suppose now that we are interested in
evaluating the probability of measuring particular values of an
\textit{observable}~$O$ of the final time configuration $u(\cdot,
t=0)$ in a subset $A \subset \RR$. Then, in the small noise limit
$\eps \to 0$, the conditional path density and the probability will be
dominated by the least unlikely path, in the sense that
\begin{align}
  \label{eq:inst-probability}
P(O[u(\cdot, 0)] \in A) = \left< 1_{\{O[u(\cdot, 0)] \in A\}} \right>
\overset{\eps \to 0}{\asymp} \exp\bigg\{- \frac{1}{\eps} \;
\inf_{\substack{\tilde{u}(\cdot, -T) = u_0\\O[\tilde{u}(\cdot, 0)] \in A}}
S[\tilde{u}]\bigg\}\,,
\end{align}
where $1_{\{\cdot \}}$ denotes the indicator function and ``$\asymp$''
stands for log-asymptotic equivalence (i.e.~the logarithms of both
sides are equal up to first order~\cite{touchette:2009}). This follows
formally by applying Laplace's method to the path
integral~(\ref{eq:path-integral}), or more rigorously by
Freidlin-Wentzell theory~\cite{freidlin-wentzell:2012}. We denote by
$u_{\text{I}}$ the field configuration for which the functional $S$
attains its global minimum for the given boundary conditions, i.e.,
$u_{\text{I}}$ solves the following minimisation problem:
\begin{align}
  \begin{cases}
  \label{eq:instanton-problem}
  \min_u S[u]\,, \\
  \text{subject to  }  u(\cdot, -T) = u_0\,, \\
  \qquad \qquad\:\: O[u(\cdot, 0)] \in A\,.
  \end{cases}
\end{align}
We call $u_{\text{I}}$ the \textit{instanton} and
$S_{\text{I}} = S[u_{\text{I}}]$ the instanton action, and can thus gain access to
limiting estimates of probabilities or probability density functions
(PDFs) in the small noise limit $\eps\to0$ by solving the
\textit{deterministic} optimisation problem of finding $u_{\text{I}}$
via~(\ref{eq:instanton-problem}). For the estimation of PDFs
$\rho_O(a)$, the target set is $A = [a, a + \mathrm{d}a]$ and
hence the optimal field configuration is sought by minimising
the action functional $S[u]$ subject to the constraint
$O[u(\cdot, 0)] = a$, which is equivalent to maximizing its
probability in path space.
Introducing $p = \chi^{-1} * [\partial_t{u} + N(u)]$, we can reformulate \eqref{eq:instanton-problem} as a the minimisation problem with respect to $p$ given by
\begin{align}
\begin{cases}
\min_p S[p] = \min_p \; \frac{1}{2} \int_{-T}^0
\left(p, \chi * p \right)_{L^2(\Omega,\RR^3)}
\dd t\,,\\[.1cm]
\text{subject to  } \partial_t{u} + N(u) = \chi * p\,,\\
\qquad \qquad \:\: u(\cdot, -T) = u_0\,,\\
\qquad \qquad \:\: O[u(\cdot, 0)] = a\,,
\end{cases}
\label{eq:min-constrained}
\end{align}
with $u = u[p]$ being a function of the control $p$ that is given by solving the PDE $\partial_t{u} + N(u) = \chi * p$, $u(\cdot, -T) = u_0$, forward in time.

We denote by $p_{\mathrm{I}}$ the optimal control and by $u_\text{I} = u[p_{\text{I}}]$ the associated optimal state.
Then, the necessary optimality conditions for \eqref{eq:min-constrained}
(derived by using a formal Lagrange approach and eliminating the adjoint state variable afterwards, compare section~\ref{appendix:grad-lbfgs}~\ref{appendix:subsec-inst-eq} in the supplemental material) yield the instanton equations
\begin{align}
\begin{cases}
\partial_t{u}_{\mathrm{I}} + N(u_{\mathrm{I}}) = \chi * p_{\mathrm{I}}\,,\\
\partial_t{p}_{\mathrm{I}} -(\nabla N(u_{\mathrm{I}}))^\top p_{\mathrm{I}} = 0\,,\\
u_{\mathrm{I}}(\cdot, -T) = u_0\,, \quad O[u_{\mathrm{I}}(\cdot, 0)] = a\,,\\
p_{\mathrm{I}}(\cdot, 0)  = -  \left. \frac{\delta O}{\delta u}^\top
\right|_{u_{\mathrm{I}}(0)} {\cal F}_{\mathrm{I}}\,.
\end{cases}
\label{eq:inst-general}
\end{align}
Here, ${\cal F}_{\mathrm{I}}$ is a Lagrange multiplier to enforce the
final time constraint.

Note that we started our considerations by expressing the probability
of an event via a path integral. The final object
we obtain though, namely the instanton, is interesting in its own
right in that it is exactly the \emph{most likely} realisation of the
outcome we set out to observe, regardless of whether it indeed
represents the \emph{typical} realisation of that outcome. The crucial
subtlety here is that for a common event there are usually a multitude
of possible histories for its creation, while an extreme outlier event
is usually driven by a very specific and reproducible chain of
events. The \emph{average} field configuration realising a moderate
vorticity, say, will in general be very different from its \emph{most
likely} configuration, and in fact is rather meaningless, as it
averages over many different and unrelated physical mechanisms. For
\emph{extreme} events, on the other hand, the two notions coincide,
and the most likely conditioned configuration precisely
corresponds to the conditioned field average.

The connection to Freidlin-Wentzell
theory~\cite{freidlin-wentzell:2012} and large deviation theory rare events
algorithms~\cite{grafke-vanden-eijnden:2019} allows us to make this
notion rather precise: The large deviation limit in the setup that was
outlined above is correct in the small noise limit. Through a
suitable rescaling of~(\ref{eq:nse-initial}), this limit is, in the
first instance, equivalent to the low Reynolds number limit for the
NSE: Non-dimensionalising all variables via $\tilde{x} = x/x_0$,
$\tilde{t} = t/t_0$, $\tilde{u} = u t_0/x_0$, $\tilde{P} = P
t_0^2/x_0^2$ and $\tilde{\eta} = \eta t_0^{1/2}/\chi_0^{1/2}$ and
choosing $t_0 = x_0^2/\nu$ yields
\begin{align}
\begin{cases}
    \partial_t u + (u\cdot\nabla)u = -\nabla P + \Delta u + \sqrt{\eps}\eta\,,\\
    \nabla\cdot u = 0\,,\\
    u(\cdot, -T) = u_0\,,
  \end{cases}
  \label{eq:nse-rescaled}
\end{align}
in the new variables. Here, $\eps = \chi_0 x_0^4 \nu^{-3} =
\text{Re}^3$ if $x_0$ is taken to be the characteristic length scale
of the forcing and the characteristic velocity $u_0$ for the Reynolds
number $\text{Re} = u_0 x_0 / \nu$ is chosen as $u_0 = (\chi_0
x_0)^{1/3}$. This shows that as $\text{Re} \to 0$, the instanton
prediction for quantities such as $\rho_O(a)$ will become
asymptotically exact for the full range of the PDF. In contrast to
this setup, we are interested in flows at a given and possibly large
Reynolds number. This can be achieved by realising that the small
noise (small $\text{Re}$) limit can be exchanged for an extreme event
limit (see remark 1 in~\cite{schorlepp-grafke-grauer:2021}): If the
length and time scales are chosen such that $\eps = \chi_0/(\nu a_0^2)$, and we focus on an event with $|O[u(\cdot, 0)]| = a_0 \gg \sqrt{\chi_0 / \nu}$
(for an observable with dimension velocity over length), for a given
Reynolds number, the instanton estimate for the typical event itself
and its probability will be accurate for sufficiently large~$a_0$ or
sufficiently extreme events. For high $\text{Re}$, these observables
must take very extreme values for the scaling limit to apply, making
it very hard to observe in DNS. As a consequence and as we will
confirm numerically in section~\ref{sec:results}, the instanton
scaling is readily reached for small Reynolds numbers, while it is
entirely out of reach of direct sampling for high $\text{Re}$, because
we are probing the tail scaling for extremely unlikely events.
This associates instantons with structures deep within the dissipation
range. We remark that the formation of these nearly singular dissipative
structures (see sections~\ref{sec:results}~\ref{sec:extr-vort-events}
and \ref{sec:results}~\ref{sec:extr-stra-events}) might be the cause
of the dissipation anomaly~\cite{duchon-robert:1999}.

\subsection{Instanton equations for Navier-Stokes with axisymmetric observables}
\label{sec:instanton-eq-nse-obsv}

For the NSE~(\ref{eq:nse-rescaled}), the instanton equations~(\ref{eq:inst-general})
can be written as
\begin{align}
\begin{cases}
\partial_t u_{\mathrm{I}} + \PP \left[(u_{\mathrm{I}}
\cdot\nabla) u_{\mathrm{I}} \right] - \Delta u_{\mathrm{I}} = \chi * p_{\mathrm{I}}\,,\\
\partial_t p_{\mathrm{I}} + \PP \left[(u_{\mathrm{I}} \cdot\nabla) p_{\mathrm{I}} +
(\nabla p_{\mathrm{I}})^\top u_{\mathrm{I}}\right] + \Delta p_{\mathrm{I}} = 0\,,\\
u_{\mathrm{I}}(\cdot, -T) = u_0\,, \quad O[u_{\mathrm{I}}(\cdot, 0)] = a\,,\\
p_{\mathrm{I}}(\cdot, 0) = - \PP\left[ \left. \frac{\delta O}{\delta u}^\top
\right|_{u_{\mathrm{I}}(0)} {\cal F}_{\mathrm{I}}\right]
\end{cases}
\label{eq:inst-nse-proj}
\end{align}
in coordinate-free form with $\nabla \cdot u_{\mathrm{I}}
= \nabla \cdot p_{\mathrm{I}} = 0$. Here, the Leray projection
$\PP = \text{Id} - \nabla \Delta^{-1} \nabla \cdot$ onto the
divergence-free part of a vector field~\cite{majda-bertozzi:2001} has been
introduced in order to eliminate the pressure from the equations of motion
and conveniently handle the incompressibility constraint within
the general framework that has been
presented in the previous section. A detailed derivation
of~(\ref{eq:inst-nse-proj}) is carried out in section~\ref{appendix:3d-nse-impl}~\ref{appendix:subsec-projection} of the supplemental
material.

We note that it is very challenging to make mathematically rigorous statements about the (unique) solvability of the corresponding minimisation problem \eqref{eq:min-constrained} for the NSE, as well as about the question of local versus global solutions. We do not attempt to do this here. Instead, for assessing the validity of our results, we rely on the following observations: i) For the simplified case of the heat equation, the equation for $p$ in~(\ref{eq:inst-nse-proj}) is independent of $u$ and the instanton equations can be solved directly without an iterative procedure and thus that problem has a unique global minimiser; ii) For the NSE, we can get a good indication whether a numerically found solution to the minimisation problem is indeed globally optimal by comparing with the PDF as obtained via DNS; iii) For the NSE, we numerically validated our claim concerning the global optimality by restarting our optimisation algorithms at various points. Within this reasoning, we will assume in the numerical results that we have found globally optimal solutions.

We are interested in extreme events for two distinct one-dimensional
observables: the vorticity $O_1[u(\cdot, 0)] = (\nabla \times
u)_z(0,0) = \omega_z(0,0)$, and the strain $O_2[u(\cdot, 0)] =
\partial_z u_z(0,0)$.  These observables correspond to the transversal
and longitudinal components, respectively, of the velocity gradient
tensor. Due to statistical isotropy and spatial stationarity, we are free to
choose the respective $z$ components as observables, as well as the origin
$x=0$ as the arbitrary point where the observables are evaluated.

Both observables naturally define a distinguished axis, around which
the problem is rotationally symmetric. In particular, not only are the
NSE rotationally symmetric, but also their corresponding action
\emph{including the conditioning on the observable} is invariant under
rotation around this axis.  It is therefore intuitive to search for a
rotationally symmetric minimiser, and this is indeed also the nature
of the structures that immediately come to mind for vorticity and
strain: Strong vorticity will be observed at the core of a
particularly strong vortex filament, while large strain occurs at
points such as the centre of the collision of two vortex rings. Of
course it is not necessarily true that a rotationally symmetric
optimisation problem has a rotationally symmetric minimiser.

Because of this fact, we set out to search for multiple, possibly
distinct minimisers of the action: One for which we artificially enforce
rotational symmetry, and potentially others for which no symmetry is
enforced. The former case reduces the problem to (2+1) dimensions in
$(r,z,t)$ for cylindrical coordinates $(r,\theta, z)$ in space. In
this coordinate system, and using the vorticity-streamfunction
formulation for axisymmetric flows~\cite{batchelor:2000}, the
\emph{axisymmetric} instanton equations are

\begin{align}
\begin{cases}
D_t u_\theta + \frac{1}{r} u_r u_\theta - L u_\theta = [\chi * p]_\theta\,,\\
D_t \omega_\theta - \frac{1}{r} u_r \omega_\theta - \frac{1}{r} \partial_z (u_\theta^2) - L \omega_\theta = [(\nabla \times \chi) * p]_\theta\,,\\
D_t p_\theta + \frac{1}{r} \left(2 u_\theta p_r - u_r p_\theta \right) + L p_\theta = 0\,,\\
D_t \sigma_\theta - \frac{1}{r} \partial_z (u_\theta p_\theta) + \partial_z u_\theta \partial_r p_\theta - \partial_r u_\theta \partial_z p_\theta \\
\quad +2 (\omega_\theta + 2 \partial_r u_z) \partial_r p_r + \frac{2}{r} \partial_r u_z p_r + \left(2 \partial_z u_z + \frac{u_r}{r} \right) \left(2 \partial_z p_r - \sigma_\theta \right) + L \sigma_\theta = 0\,,
\label{eq:inst-cylin}
\end{cases}
\end{align}
where $D_t = \partial_t + u_r \partial_r + u_z \partial_z$ is the axisymmetric
convective derivative, $L = \frac{1}{r} \partial_r(r \partial_r)
- \frac{1}{r^2} + \partial_{zz}$
is an elliptic operator stemming from the vector Laplacian in cylindrical
coordinates, and $\sigma = \nabla \times p$ is the vorticity of the adjoint field.
In this formulation, the $r$ and $z$ components of the fields are reconstructed by
solving $L \psi = - \omega_\theta$ for the streamfunction $\psi$ and computing
$u_r = - \partial_z \psi$ and $u_z = \frac{1}{r} \partial_r(r \psi)$.

The derivation of~(\ref{eq:inst-cylin}), as well as the spatio-temporal
boundary conditions of the axisymmetric instanton fields, can be found in section~\ref{appendix:2d-axi-impl}~\ref{appendix:subsec-axinst-bc} of the supplemental material.

\subsection{Numerical procedure}
\label{sec:numerical-procedure}
We consider the problem of minimising the action functional $S[u]$ given by \eqref{eq:om-action} subject to a final time constraint $O[u(\cdot,0)] = a \in \RR$. Here, we briefly outline the
numerical procedure that we use to compute axisymmetric and fully three-dimensional solutions.

We interpret the minimisation problem within the framework of PDE-constrained optimal control (see e.g.~\cite{troltzsch:2010,herzog-kunisch:2010}): We introduce the
control variable $p$ as discussed above and consider the optimisation problem \eqref{eq:min-constrained}.
This is an optimal control problem with distributed control as the control $p$ enters the PDE on the right hand side as a source term. 
The velocity field $u=u[p]$ corresponds to the state variable. By treating $u$ as a function of $p$, we can follow the so-called reduced approach in optimal control theory and
view the optimal control problem as a problem of the argument $p$ only.
We can recast \eqref{eq:min-constrained} into a sequence of unconstrained optimisation problems using the augmented Lagrangian
method~\cite{hestenes:1969}:
For a sequence of positive penalty parameters $(\mu^{(m)})$ with
$\mu^{(m)} \to \infty$ we minimise
\begin{align}
L_{\text{A}}[p, {\cal F}, \mu] =S[p] + {\cal F} (O[u[p](\cdot, 0)] - a) +
\frac{\mu}{2} (O[u[p](\cdot, 0)] - a)^2\,,
\label{eq:augmented-lagrangian}
\end{align}
while updating the Lagrange multiplier ${\cal F}$ via ${\cal F}^{(m + 1)} =
  {\cal F}^{(m)} + \mu^{(m)} \left(O[u[p^{(m)}](\cdot, 0)] - a \right)$.

  In other words, for each penalty parameter $\mu^{(m)}$, we need to solve a minimization problem, which in turn requires an iterative scheme. The computational costs can be reduced by using warm starts.
This procedure allows us to compute instantons for a specified observable value $a$.  
This is in contrast to the optimisation approach by Chernykh and
Stepanov~\cite{chernykh-stepanov:2001} and others. There, the instanton equations, compare \eqref{eq:inst-general} for a generic SPDE, are solved by an iterative procedure during which the Lagrange multiplier ${\cal F}$ is kept fixed and the value of $a$ is allowed to change. This again produces a solution to the instanton equations, including a matching pair $({\cal F}_\text{I},a)$, but the value of $a$ is not known a priori. 
This practical approach is convenient and computationally cheaper if there is a bijective map between ${\cal F}$ and $a$ and one is interested in solving the instanton equations over a wide range of values of $a$. 
Our approach is more general and to be preferred if  i) there
are multiple local minimisers, and the map ${\cal F} \mapsto a$ becomes
multivalued, and ii) there are observable regions where the action fails to
be convex and the ${\cal F}$-$a$-duality breaks down~\cite{alqahtani-grafke:2021}.

To minimise \eqref{eq:augmented-lagrangian} for a given value $\mu^{(m)}$, we employ gradient based methods.
As an improvement over a simple
gradient descent (which, preconditioned with~$\chi^{-1}$, reduces to an
iterative, fixed-point like solution of the instanton equations),
we use the L-BFGS algorithm~(see e.g.~\cite{nocedal-wright:2006}).
This significantly speeds up the computation for
the fully three-dimensional instantons.
The L-BFGS scheme is a limited-memory variant of one of the most popoular quasi-Newton schemes, the BFGS scheme, named after Broyden, Fletcher, Goldfarb and Shanno.
Quasi-Newton schemes only require gradient information (in contrast to the second-order derivative information needed for Newton) and typically show super-linear convergence (whereas the gradient scheme only converges linearly with rates that are often very close to 1 for ill-conditioned problems).
Appropriate step sizes for the
optimisation algorithm are determined by an Armijo line search using backtracking
\cite{nocedal-wright:2006}. This very popular condition guarantees sufficient decrease that is proportional to the step length. For the evaluation of the gradient, we use an adjoint approach: the gradient is given as $\delta L_{\text{A}}/\delta p = \chi * (p -z)$,
where the adjoint state $z$ solves the backward
equation $\partial_t{z} - (\nabla N(u[p]))^\top z = 0$ with final condition
$z(\cdot, 0) = -(\delta O/\delta u\vert_{u[p](\cdot,0)})^\top
({\cal F} + \mu (O[u[p](\cdot, 0)] - a))$. Thus, each gradient evaluation requires to solve a PDE forward in time to determine~$u[p]$ and then backwards to compute~$z$. All of this is described in detail in section~\ref{appendix:grad-lbfgs} in the supplemental material.

We use two different flow solvers within the described optimisation
framework: a (2+1)-dimensional axisymmetric code as well as a
(3+1)-dimensional code for the full problem. The (2+1)-dimensional
code is necessary to compute solutions of the minimisation problem
under the additional constraint of preserving axisymmetry.  For the
(3+1)-dimensional code, the rotationally symmetric instanton
eventually ceases to be a local minimiser of the action as there are unstable
directions that break symmetry. Symmetrisation stabilises the
configurations and allows us to get access to the associated
action. In other words, after symmetry breaking, the axisymmetric
configuration ceases to be a minimiser of the full optimisation problem,
but remains a (local) minimiser of the axisymmetric optimisation
problem. The axisymmetric code is based on \cite{grauer-sideris:1991}:
We use a Leapfrog scheme in time and symmetric second order finite
differences on a regular $r$-$z$-grid in space, with a resolution of
$n_t = 1024$ and $n_r = n_z = 256$. The diffusion term is discretised
semi-implicitly to avoid a severe CFL constraint.  Consequently, in
each time step, we need to solve a Helmholtz-like equation to update
the fields, for which we use a multigrid algorithm (see
e.g.~\cite{trottenberg-etal:2000}).  The polar convolutions
with~$\chi$ are evaluated by means of fast Hankel
transforms~\cite{fisk-johnson:1987,melchert-wollweber-roth:2018}.

The full (3+1)-dimensional flow solver uses a pseudo-spectral method in space
and the Heun scheme in time, with an integrating factor for the diffusion term.
Thus, we again avoid a strict CFL constraint. We run a resolution
of $n_t = 512$ and $n_x = n_y = n_z = 128$. For speed up, we implemented this on
a GPU using the CUDA API. To fit a full (3+1)-dimensional
optimisation problem on a single GPU, memory reduction techniques as described
in~\cite{grafke-grauer-schindel:2015} were necessary.

\section{Results}
\label{sec:results}

In the following, we show the outcome of our numerical computations,
beginning with the instanton configurations before and after symmetry
breaking for both vorticity and strain. We then discuss implications
on the tail scaling of the PDFs, in particular for
large positive strain.

\subsection{Extreme vorticity events}
\label{sec:extr-vort-events}

\begin{figure}
\centering
\includegraphics[width = \textwidth]{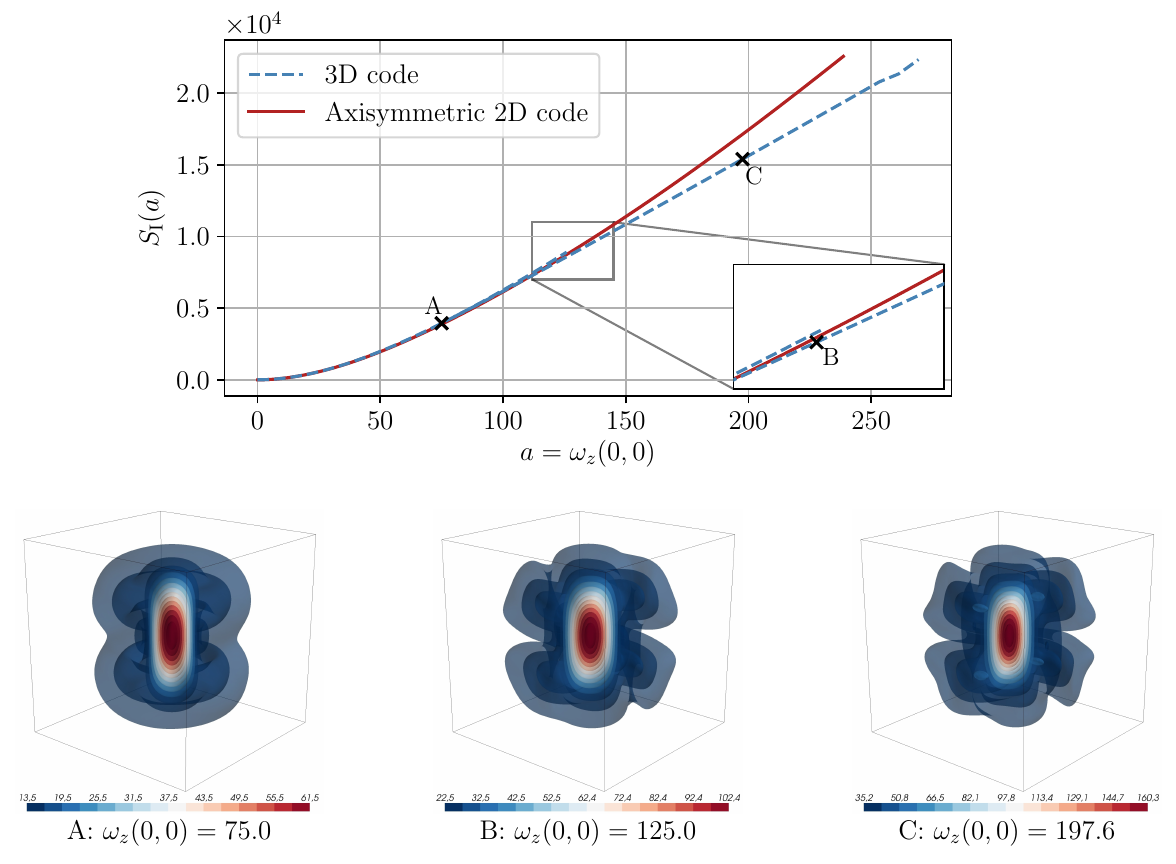}
\caption{Results of the full 3D and axisymmetric instanton
  computations for the vorticity observable $\omega_z(0,0)$. The plot
  in the top row shows the action $S_{\text{I}}(a)$ at all critical
  points of the action that were found in our numerical experiments
  for different values of the final-time constraint $\omega_z(0,0) =
  a$. The bottom row shows isosurfaces of the vorticity of the
  final-time configuration of the obtained instanton fields for
  different observable values as indicated in the top
  plot. Qualitatively, the field configurations which we observe are
  vortex tubes in all cases. However, the 3D computations
  show that a second branch that breaks full rotational symmetry and
  reduces to reflection symmetry dominates the fully symmetric
  branch in probability and splits off at $a_{\text{c}} \approx 85$.}
\label{fig:vorticity-result}
\end{figure}

Selecting $\omega_z(0,0)=a$ as our observable, we use the above
formalism to numerically solve the optimisation problem
(\ref{eq:instanton-problem}). The result is the most likely
configuration to realise an extreme vorticity outcome at final
time. Note that this computation is independent of the choice of the
Reynolds number. The Reynolds number, or equivalently $\eps$, only
determines whether a chosen observable $a$ is rare, and thus whether
the instanton formalism has any relevance for events of this size. As
shown in figure~\ref{fig:vorticity-result}, the most likely
configuration to realise an extreme vorticity corresponds to a vortex
filament with an added swirl component. We first show, in the top row
of figure~\ref{fig:vorticity-result}, how the full 3D and the
axisymmetric code find the same minimiser for low values of $a$, but
find different minimisers for high values. Configuration A, at $a=75.0$,
is still in the regime where the global minimiser is rotationally
symmetric. At configuration B, for $a=125.0$, the symmetry-broken
branch has already appeared, but is still very close to the symmetric
one. Configuration C, at $a=197.6$, is in a regime where the symmetry
broken minimiser clearly dominates the symmetric minimiser. The
asymmetric minimising configurations correspond to vortex tubes with a
symmetry-breaking helical vortex structure around it that displays only
reflection symmetry instead of full axial symmetry. Due to the symmetry
of the minimisation problem under reflection with respect to
the $z = 0$ plane, the behaviour is identical for
negative $a$, with a mere sign-flip in $\omega$ (not shown).

Note that around the point of symmetry breaking, the full 3D code
picks up both the symmetric and asymmetric minimisers until the
symmetric configuration eventually becomes unstable, as indicated by
the two blue dashed lines in the inset of
figure~\ref{fig:vorticity-result} (top), where the upper line
corresponds to the rotationally symmetric local minimiser of the full
3D code. There is a small difference between the
rotationally symmetric minimiser of the full 3D code, and the same
minimiser for the axisymmetric code, which is the result of numerical
differences in the integration schemes and coordinate systems.

\begin{figure}
\centering
\includegraphics[width = \textwidth]{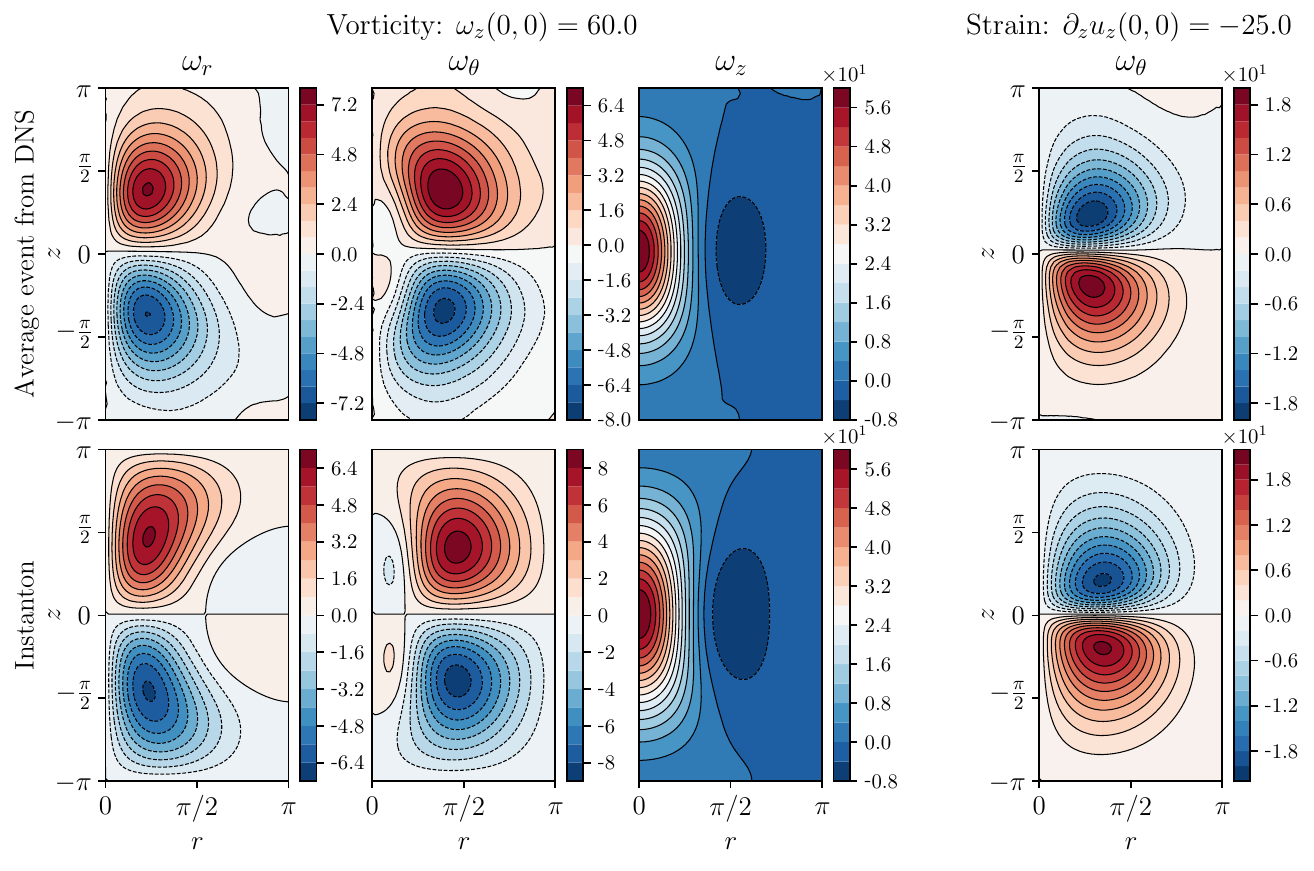}
\caption{Comparison of the final-time field configuration of
  \textit{axisymmetric} vorticity and strain instantons (bottom row)
  to conditional averages of DNS data for the same prescribed
  observable values at the origin as the instanton fields (top
  row). The left three columns show all components of the vorticity in
  cylindrical coordinates for an event with a prescribed value
  of~$\omega_z(x = 0, t = 0) = 60.0$ at the origin. The rightmost column
  only shows the $\theta$ component of the vorticity of an event
  with~$\partial_z u_z(x = 0, t = 0) = -25.0$ since the~$\omega_r$
  and~$\omega_z$ components are negligibly small. The conditional
  averages of the DNS data include an angle averaging procedure in
  $\theta$, and events with suitable observable values at $x \neq 0$
  were shifted onto the origin. For the displayed vorticity event,
  approximately $8.4 \cdot 10^3$ single events as obtained from DNS
  of~(\ref{eq:nse-rescaled}) with a forcing strength of~$\eps = 250$
  were averaged, whereas the strain event is an average of
  approximately $5.1 \cdot 10^3$ events in the same data set.}
\label{fig:filter-comparison}
\end{figure}

We can compare the instanton configuration against structures observed
in DNS, conditioned on observing an extreme vorticity
event~\cite{novikov:1993}.  The result of this ``filtering''
procedure~\cite{grafke-grauer-schaefer:2013} is shown in
figure~\ref{fig:filter-comparison} (left three columns) for the
axisymmetric configuration only. Concretely, this compares an
instanton for $\omega_z=a=60.0$, which would be located left of
configuration A in figure~\ref{fig:vorticity-result}, in cylindrical
coordinates, against the conditional average of DNS data at
$\eps=250$, conditioned on $\omega_z=60.0$. To compute this average,
we integrate $10^4$ independent realisations of the
3D~NSE~(\ref{eq:nse-rescaled}) on $\Omega = [0, 2 \pi]^3$ 
for a total time of $T=1$ used in all computations throughout
this paper (which is much larger than the large eddy turnover time
$T_{\text{LET}} \approx 0.1$ for this $\eps$).  Exploiting the
statistical isotropy and homogeneity of the system in order
to increase the sample size, we analyse the final
field configuration for events with $\big|\lVert\omega(x)\rVert - a\big|/a <
0.01$, and then rotate and translate the coordinate system so that the
event is located at $x=(0,0,0)$ and points in $z$-direction. We
average $8.4 \cdot 10^3$ such events, including averaging in $\theta$
for each individual event, to obtain the results of
figure~\ref{fig:filter-comparison} (top row). The conditional
average obtained in this way agrees excellently with the instanton
event for the same vorticity, demonstrating that for this $\text{Re}$
the most likely and the average configuration realising
$\omega_z=60.0$ are identical, and we are indeed in the large
deviation limit.

\subsection{Extreme strain events}
\label{sec:extr-stra-events}

\begin{figure}
\centering
\includegraphics[width = \textwidth]{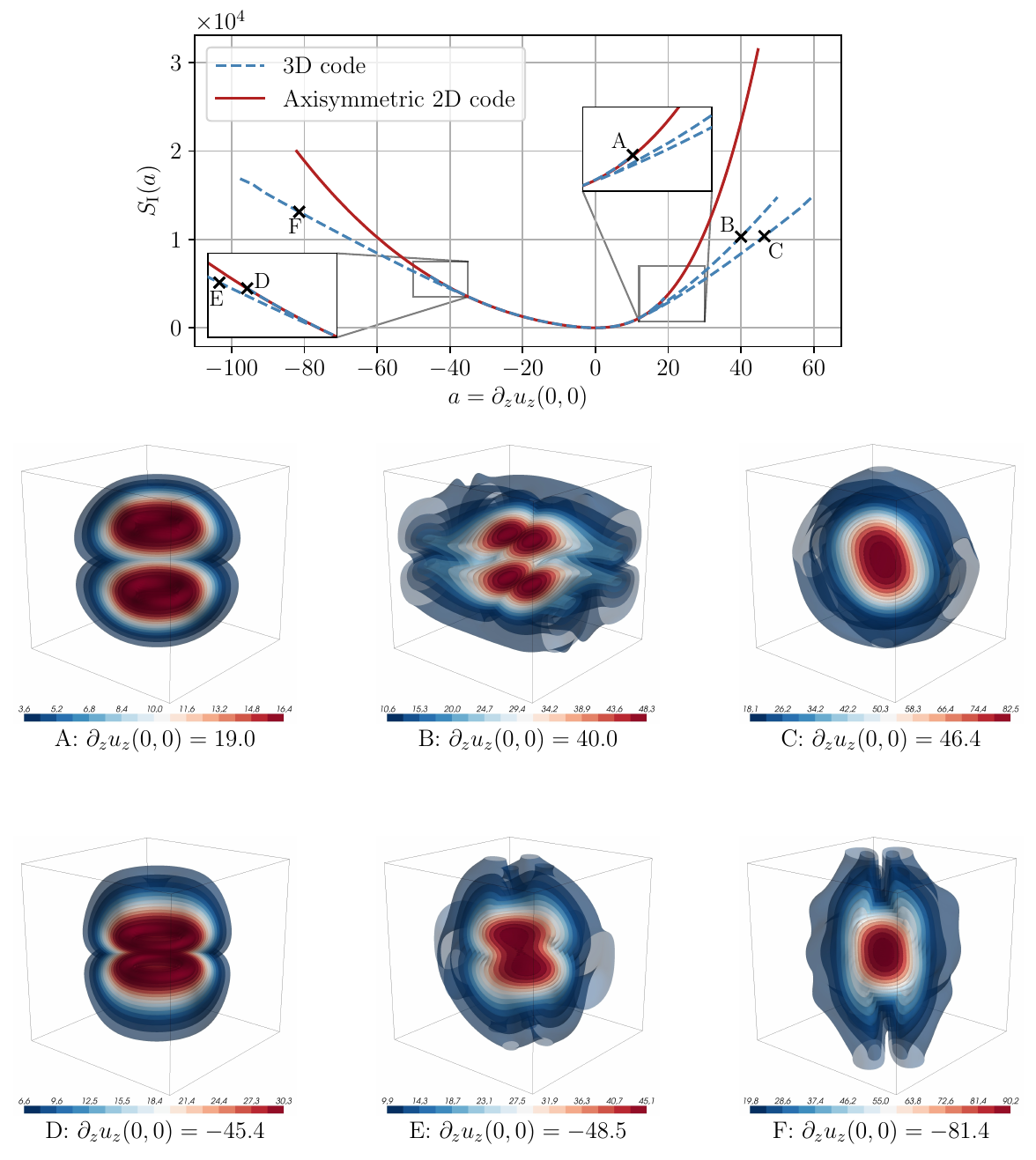}
\caption{Results of the axisymmetric and full 3D instanton
  computations for the strain observable $\partial_z u_z(0,0)$. As in
  Figure~\ref{fig:vorticity-result}, the top plot shows the action at
  all critical points that were found numerically for different
  observable values, and the two bottom rows show isosurfaces of the
  vorticity of the final-time configuration of the indicated instanton
  fields.  Note that, contrary to the vorticity instanton, we find a
  qualitative difference between the rotationally symmetric strain
  instanton consisting of two counter-rotating vortex rings (A and D)
  and a dominant, symmetry-breaking instanton branch that consists of
  thin vortex sheets (C, E and F). Furthermore, for large positive
  strain, we find a third, subdominant branch with a quadrupole-like
  symmetry (B).}
\label{fig:strain-result}
\end{figure}

Performing the same procedure for the strain observable, $\partial_z
u_z(0,0) = a$, we obtain a richer set of outcomes. For the strain, positive
and negative observables have different phenomenologies caused by the
advection term (see e.g.~\cite{wilczek-meneveau:2014}),
but both eventually undergo symmetry
breaking. As visible in figure~\ref{fig:strain-result} (top), the
earliest and most dramatic symmetry breaking is observed for the
positive tail of the strain, where an asymmetric branch splits off
already at $a_{\text{c}} \approx 14$. Here, the
symmetric configuration A, consisting of two counter-rotating, contracting
vortex rings, transitions for higher $a$ into an
asymmetric sheet/pancake like structure C. We additionally observe a
further subdominant symmetry-breaking branch of quadrupole-like
configurations B. It is of course difficult to exclude the existence
of further subdominant local minimisers, but our numerical experiments where
we started the optimisation algorithm either at a random initial condition
for the control or at perturbed solutions of previous problems did not
show indications of further branches in the considered observable range.

The negative tail has qualitatively similar behaviour at different
values $a$: The symmetric configuration D, corresponding to two colliding
vortex rings with opposite orientation, breaks away at $a_{\text{c}}\approx-38$
into more complicated and asymmetric vortex sheet configurations E and F.

The vortex ring structures that we encountered are ``trivial'' solutions of
the instanton equations~(\ref{eq:inst-cylin}) in cylindrical coordinates in
the sense that they satisfy $\omega_r = \omega_z \equiv 0$ which does not
yield the global minimum of the full action at large observable values.
Interestingly, field configurations of this type have previously been found
as maximisers of the enstrophy growth rate in~\cite{lu-doering:2008}.
On the other hand, sheet-like structures, as in the symmetry breaking case, have
been observed as the most intense dissipative structures already in
\cite{moisy-jimenez:2004} and in recent spectral simulations using $8192^3$ grid
points~\cite{iyer-schumacher-sreenivasan:2019}.

We further compare the rotationally symmetric strain instanton to the
conditional average from DNS in figure~\ref{fig:filter-comparison}
(rightmost column). Here, with the data set and procedure as for the
vorticity, we only compare the $\omega_\theta$-component in
cylindrical coordinates, with excellent agreement between the
minimiser and the observed conditionally average event realising the
strain value of $a=\partial_z u_z(0,0) = -25.0$. For the vortex ring
configuration, all other components are negligibly small ($\approx
10^{-4}$ for the 3D instanton code due to numerical noise, $\approx
10^{-1}$ due to statistical noise in the DNS average). We do not
compare the symmetry broken instantons to conditional averages, since
this would require a much larger data set, where each event, instead of
being averaged over~$\theta$, is additionally aligned in angular
direction, using e.g.~the eigenvectors of the velocity gradient
tensor.

\subsection{Extreme event probabilities}
\label{sec:extr-event-prob}

\begin{figure}
\centering
\includegraphics[width = \textwidth]{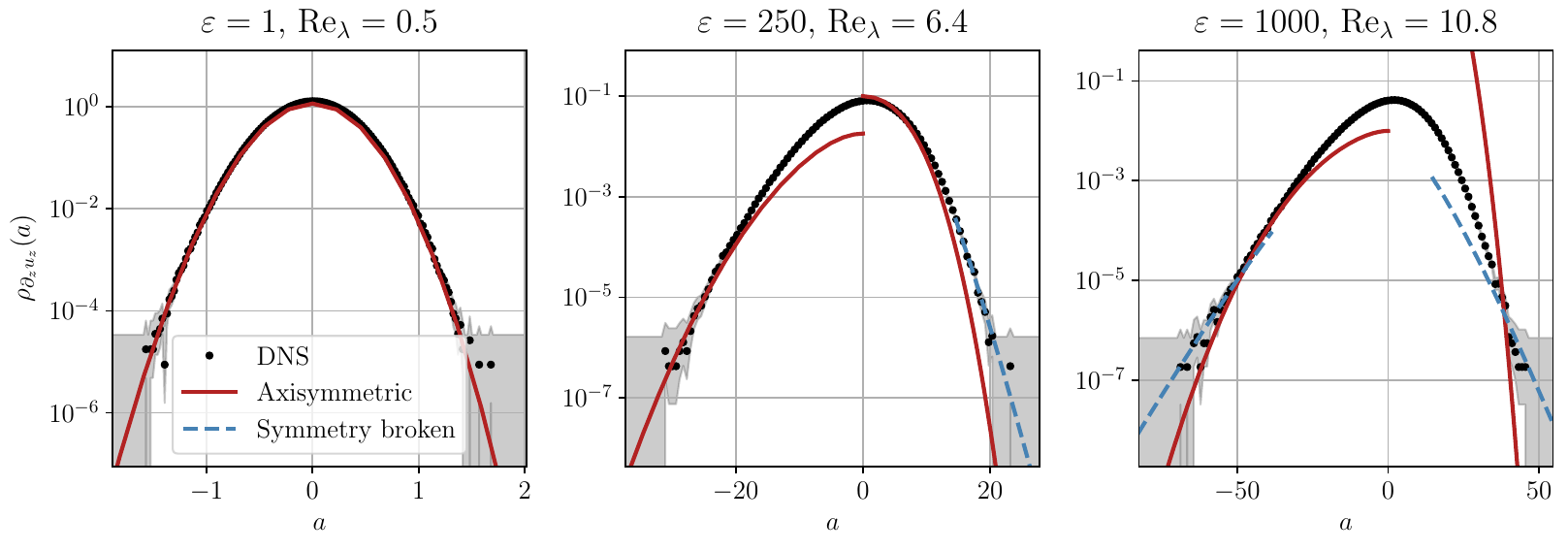}
\caption{Comparison of the instanton prediction $\propto \exp\{-
  \eps^{-1} S_{\text{I}}(a)\}$ for the strain PDF~$\rho_{\partial_z
    u_z}$ to DNS data at different forcing strengths~$\eps$ or
  Taylor-Reynolds numbers $\text{Re}_\lambda$. The dots show the DNS
  histogram, with a $95\%$ Wilson score
  interval~\cite{wilson:1927,brown-cai-dasgupta:2001} shaded in grey. The
  solid lines show the PDF prediction as obtained from the
  axisymmetric instanton configurations, whereas the dashed lines show
  the PDF prediction based on the \textit{lowest} symmetry broken
  branch of the instanton action. Note that we are free to
  shift all individual branches arbitrarily and independently in
  the vertical direction in the semi-logarithmic PDF plot since
  we are only interested in asymptotic scaling estimates.
  Observe in particular that the axisymmetric strain instanton
  clearly underestimates the right tail
  even at the small Reynolds numbers considered here.}
\label{fig:strain-pdfs}
\end{figure}

The derivation of the instanton formalism in
section~\ref{sec:instantons-3d-navier}, and in particular
equation~(\ref{eq:inst-probability}) make obvious that the instanton
not only represents the most likely extreme event, but further allows
us to estimate its probability, which scales exponentially with the
instanton action. In this section, we compare this prediction for the
exponential scaling of the tail with PDFs obtained from DNS, and in
particular demonstrate how the symmetry-broken instanton predicts the
correct tail scaling for the PDFs, while the axisymmetric instantons
dramatically underestimate the likelihood of large strain events. We
concentrate on the positive tail of the strain observable in
particular, since there the symmetry is broken the earliest and the
difference in slope is the clearest.

Figure~\ref{fig:strain-pdfs} shows this comparison for three different
values $\eps\in\{1, 250, 1000\}$, corresponding to three different
Taylor-Reynolds numbers $\text{Re}_\lambda = \sqrt{15 \text{Re}} \in
\{0.5, 6.4, 10.8\}$ (where $\text{Re}$ was determined from the root
mean square velocity and integral scale of the data). Note that even
the highest $\text{Re}$ is still comparably low. This is because, as
argued in
section~\ref{sec:instantons-3d-navier}~\ref{sec:acti-minim-inst}, for
higher $\text{Re}$ the instanton is so far in the tail that it cannot
be detected in DNS.

For each $\eps$, we performed $10^4$
pseudo-spectral simulations of the 3D~NSE~(\ref{eq:nse-rescaled}) at a spatial
resolution of $128^3$ starting from $u_0 = 0$ at~$T = -1$ until~$t =
0$. The final time configurations were subsampled according to the
estimated approximate correlation length $\lambda_{\partial_z u_z} =
0.8$ of the observable, and the shaded area indicates a $95\%$ Wilson
score interval~\cite{wilson:1927,brown-cai-dasgupta:2001} for the
PDF estimate based on the DNS data. For the
lowest $\text{Re}$, the data is almost Gaussian, and the instanton and
PDF agree everywhere. No symmetry breaking is observed. For the two
higher $\text{Re}$, instead, the instanton approach only captures the
tail scaling correctly, since common strain events are not
dominated by the instanton in this case. In the tails,
though, the axisymmetric instanton clearly underestimates the
probability, while the symmetry-broken instanton is in good
agreement. This is particularly clear in the right tail of the
rightmost panel of figure~\ref{fig:strain-pdfs}, where the
axisymmetric instanton depicted by the red line is far too steep to
agree anywhere with the observed tail scaling. This trend continues
in fully developed turbulence at higher Re:
The analysis of larger DNS, e.g.~in~\cite{buaria-pumir-bodenschatz-etal:2019},
shows that the strain PDF tails can in fact be described by stretched exponentials
$ \propto \exp \big\{ -c_{\pm} |a|^{\vartheta_{\pm}}\big\}$ with
exponents $\vartheta_{\pm} < 1$, whereas we find that
the exponents in both tails derived from the vortex-ring instanton increase
monotonically with $|a|$ and saturate above $\vartheta_+ = 2.5$ in the right tail
and above $\vartheta_- = 2$ in the left tail. In contrast, while the
$S_{\text{I}}(a)$-curve that we obtained for the symmetry-broken instantons
is still convex in the observable range that we were able to consider
at the given resolution, the exponents $\vartheta_{\pm}$ are
monotonically decreasing in $|a|$ for this branch and decay below $1.5$
for both positive and negative strain.

For the vorticity observable of the same DNS dataset at $\eps\in\{1, 250, 1000\}$,
we observe the same qualitative results (not shown): At the lowest Re, the
instanton again perfectly describes the PDF, whereas the range of validity of
the estimate transitions into the tails at higher Re. Here, however, because the
symmetry breaking occurs at relatively higher $a$ and leads to a less dramatic
difference in scaling, it is hard to draw as clear a conclusion as in the
strain case.

\section{Conclusion}
\label{sec:conclusion}

In this paper, we set out to numerically compute maximum likelihood
realisations of extreme vorticity and strain events in the stochastic
incompressible 3D~NSE. As an alternative and complement to direct
sampling approaches, we rephrased the problem into a deterministic
variational framework using sample path large deviation theory, which
is particularly suited for rare and extreme events. This led us to
consider a (3+1)-dimensional optimisation problem with final-time
constraints to enforce large observable values, which we were able to
solve using tools from PDE-constrained optimisation. For both
observables considered here, we observe symmetry breaking of the
minimisers: The vortex filaments that lead to large values of the
vorticity reduce from axial to reflection symmetry, and the vortex
rings that realise large strain transition to a pancake-like vortex
sheet structure. For positive strain in particular, we demonstrated
that the symmetry-broken minimiser clearly dominates the symmetric one
and can in fact be confirmed to yield the correct scaling of DNS PDFs
at suitable Re, in contrast to the axisymmetric one.

The possibility to access the most extreme events in Navier-Stokes
turbulence without sampling is attractive. Despite the fact that the
optimisation problem~(\ref{eq:instanton-problem}) to be solved is
massive, with fields of size $512\times 128^3$, and a single iteration
of the minimisation algorithm corresponding to a forward integration of
the NSE, and an equally sized backward propagation,
we show that this effort pays off for extreme outlier events:
Obtaining these same configurations traditionally necessitates either
millions of samples of the stochastic NSE (for
lower $\text{Re}$), or the regime is completely inaccessible as the
events are entirely too rare and extreme to be observed (for higher
$\text{Re}$). While one could try to formulate reduced problems in
effective coordinates, for example
as~in~\cite{kleineberg-friedrich:2013}, our approach yields the most
likely configuration without any \textit{a~priori} assumptions about
its form or physical mechanisms.

In this paper, we only considered the exponential contribution of the
minimiser for the PDF. Improved estimates are possible in principle
when taking into account the fluctuations around the instantons, as
discussed e.g.~in~\cite{schorlepp-grafke-grauer:2021}. The
computational cost of computing this fluctuation determinant is vastly
bigger than the already large problem sizes encountered in the
optimisation problem in this work. For this approach, it is further
necessary to integrate out the zero mode associated with the symmetry
breaking of the instanton. This correction to the PDF was ignored in
this paper.

It would be interesting to determine whether the viscid instanton we
discussed here has relevance to inertial range properties of turbulent
flow. One possible connection is given by the scaling of velocity
gradient moments, which, even in the low Reynolds numbers regime, link
dissipative statistics to inertial range properties via so-called
fusion rules~\cite{benzi-biferale-etal:1999,
  schumacher-sreenivasan-yakhot:2007,
  friedrich-margazoglou-etal:2018}. This possible route towards
understanding intermittency is the focus of our future work.

\acknowledgments{TS would like to thank Sebastian Gallon for useful discussions and help regarding the implementation of the CUDA flow solver and framework. TG acknowledges the support received from the EPSRC projects
  EP/T011866/1 and EP/V013319/1.}


\appendix

\section*{Appendix: Supplemental material}


\section{Introduction and Outline}
\label{appendix:constrained}

In this supplemental material, we describe in detail the numerical algorithms
that we use to minimise the Freidlin-Wentzell action
functional. This way we compute instantons or maximum likelihood
trajectories realising extreme values of vorticity and strain, and estimate the
dominant contribution of the minimisers to the vorticity and strain 
probability density functions (PDFs) for the
three-dimensional stochastic Navier-Stokes equations (NSE).

Hoping that these notes might be useful for related problems as well, we will discuss the minimisation problem
in a slightly more general setting. Nevertheless, we will also specify all necessary details for the specific implementation of the
algorithm for the 3D NSE. In these notes, we focus on expanding on section 2 of the
main paper. For a general review of the numerics of the instanton method,
we refer to~\cite{grafke-grauer-schaefer:2015, grafke-vanden-eijnden:2019}.
The main source for the optimisation techniques that we use and that we describe in the following is \cite{nocedal-wright:2006}.

\subsection{Setup of the problem}
\label{appendix:subsec-setup}

We consider a general $d$-dimensional system of stochastic partial differential
equations (SPDEs)
\begin{align}
\partial_t u(x,t) + N(u(\cdot, t))(x) = \eta(x,t),
\quad \left< \eta(x,t) \eta^\top(x',t') \right> = \eps \chi(x-x') \delta(t-t')\,,
\end{align}
for a vector field $u:\Omega \times [-T,0] \to \RR^d$ on the time
interval $[-T,0]$ with $T > 0$ and deterministic initial condition
\begin{align}
u(x,-T) = u_0(x)\,.
\end{align}
Here,~$N$ denotes a deterministic and possibly spatially non-local nonlinearity.
The spatial boundary conditions will be assumed to be periodic, such that the
domain~$\Omega = \TT^d = [0, 2 \pi]^d / \sim$ is given by a $d$-torus.
The $d$-dimensional forcing $\eta$ is assumed to be Gaussian with zero mean,
white in time, and stationary with spatial covariance matrix
function~$\chi:\Omega \to \RR^{d \times d}$, which will typically
only act on large scales in the applications considered here.
Finally, the noise strength $\eps \in \RR$ is assumed to be a small positive number.

The path integral formulation of the PDF $\rho_O$ of a general, $\RR^{d'}$-valued
observable $O[u(\cdot, 0)]$ at $a \in \RR^{d'}$ takes the form
\begin{align}
\rho_O(a) = \int Du \; \delta(u(\cdot, -T) - u_0) \delta(O[u(\cdot, 0)] - a) J[u]
\exp \left\{ - \varepsilon^{-1} S[u] \right\}\,, \label{eq:PI-general}
\end{align}
where $S$ denotes the Freidlin-Wentzell action~\cite{freidlin-wentzell:2012}
\begin{align}
S[u] = \frac{1}{2} \int_{-T}^0 \dd t \; {\cal L}(u, \partial_t u)
= \frac{1}{2} \int_{-T}^0 \dd t \; \left(\partial_t u + N(u),
\chi^{-1} * \left[\partial_t u + N(u) \right] \right)_{L^2(\Omega,\RR^d)} \geq 0\,.
\label{eq:action-om}
\end{align}
Further, $J$ is an additional term that arises from the noise-to-field
transformation $\eta \to u$ and that is independent of~$\eps$. The spatial convolution $*$ appears due to the fact that the forcing is assumed to be spatially stationary, such that
\begin{align}
\int_{\Omega} \dd^dx' \; \chi^{-1}(x,x') \eta(x') = \int_{\Omega} \dd^dx' \; \chi^{-1}(x-x') \eta(x') =(\chi^{-1} * \eta)(x)\,.
\end{align}
Here, $\chi^{-1}$
denotes the inverse operator of $\chi$ in the sense that
\begin{align}
\int_\Omega \dd^d x' \; \chi^{-1}(x-x') \chi(x' - x'') = \text{Id}\;\delta^{(d)}(x-x'')\,.
\end{align}
Hence, in Fourier space,
$\chi^{-1}$ is given by the pointwise inverse matrix of $\chi$. With $\chi$
typically acting only on large scales, as well as taking a purely solenoidal forcing
in the application that we consider, this implies
that $\chi$ will be singular. Hence, $\chi^{-1}$ will only be defined on a subspace,
and we will discuss how to deal with this singular behaviour below. 
For the definition of the action functional~(\ref{eq:action-om}), it is appropriate 
to set $S[u] = +\infty$ if $\partial_t u + N(u)$ is not in the image of the spatial convolution with $\chi$, which
corresponds to trajectories of the system that are not realisable and hence 
are assigned a probability of 0.

In an appropriate large deviation limit such as the small noise
limit $\eps \to 0$, the path integral~(\ref{eq:PI-general}) is dominated
by the contribution of the minimum action realisation of the
velocity field, or instanton, which we denote by $\uu$, defined as the solution of
\begin{align}
  \begin{cases}
  \label{eq:umin-def}
  \min_u S[u]\,, \\
  \text{subject to  }  u(\cdot, -T) = u_0\,, \\
  \qquad \qquad\:\: O[u(\cdot, 0)] = a\,.
  \end{cases}
\end{align}
Solving this constrained minimisation problem~(\ref{eq:umin-def}) for a range of values for $a$,
evaluating the action $S_{\text{I}} := S[\uu]$ at the minimum, and expanding
the path integral around this minimum directly yields the log--asymptotic
scaling of the PDF
\begin{align}
\rho_O(a) \overset{\eps \to 0}{\asymp} \exp \left\{-\eps^{-1} S_{\text{I}}(a)\right\}\,,
\label{eq:rho-log-asymp}
\end{align}
the result of which is analysed in the main text for the three-dimensional 
NSE. Sharper asymptotics in the sense of a prefactor analysis as
in~\cite{schorlepp-grafke-grauer:2021} are beyond the scope of the current work.

As discussed in the main text, the instanton is not only relevant as the
starting point of a saddle-point approximation to the path
integral~(\ref{eq:PI-general}). It is also
a meaningful object in itself as it describes the 
maximum likelihood path realising a given outcome in the following sense:
while the probability of each individual trajectory is 0, tubes of
radius~$R$ in path space that are centred around the instanton will
dominate all other tubes of radius~$R$ in probability as $R \to 0$ and $\eps \to 0$.

\subsection{Reformulation as an optimal control problem}
\label{appendix:subsec-oc}

In order to eliminate the inverse correlation function from the target
functional~(\ref{eq:action-om}) and to reduce the functional to a
quadratic form, we introduce the conjugate momentum
\begin{align}
p := \nabla_{\partial_t u} {\cal L}
= \chi^{-1} * \left[\partial_t u + N(u) \right]\,.
\end{align}
We treat $p$ as the only independent variable in the resulting minimisation problem
\begin{align}
\begin{cases}
\min_p S[p] = \min_p \; \frac{1}{2} \int_{-T}^0 \dd t
\left(p, \chi * p \right)_{L^2(\Omega,\RR^d)}\,,\\[.1cm]
\text{s.t. } O[u[p](\cdot, 0)] = a\,.
\end{cases}
\label{eq:min-constrained-supmat}
\end{align}
In the language of optimal control, the field $p$ is the (distributed) control 
of the system, and the minimisation is carried out over all such $p$ for which 
the dependent state variable $u=u[p]$, that solves the $p$-dependent PDE
\begin{align}
\begin{cases}
\partial_t u + N(u) = \chi * p\,,\\[.1cm]
u(\cdot, -T) = u_0\,,
\end{cases}
\label{eq:pde-forward}
\end{align}
satisfies the final time constraint
\begin{align}
O[u[p](\cdot, 0)] = a\,.
\label{eq:final-time-constraint}
\end{align}
Note that only $\chi * p$ enters the target 
functional and the PDE constraint. This significantly reduces 
the dimensionality of the optimisation problem if the forcing is highly degenerate, 
as is the case for our application.
This will come in very handy in the context of memory reduction techniques, which we will
discuss in section~\ref{appendix:grad-lbfgs}~\ref{appendix:redstor} below.
Furthermore, we exploit this observation to invert $\chi$ numerically:
For an arbitrary input field $q$, we define 
$\chi^{-1} * q$ by first projecting $q$ onto the image 
of $\chi$ and by then inverting $\chi$ on its domain without its kernel. 

\section{Conversion to unconstrained problems}
\label{appendix:unconstrained}

There exist various approaches for  reducing the constrained optimisation
problem~(\ref{eq:min-constrained-supmat}) to a single (or a series of) unconstrained
optimisation problem(s). We will present a selection in the following. 
Afterwards, in section \ref{appendix:grad-lbfgs}, we will discuss the solution of the resulting
unconstrained problems by means of gradient based methods.

\subsection{Method of Lagrange multipliers}
\label{appendix:subsec-lagrange-mult}

The conceptually easiest way to incorporate the final time
constraint~(\ref{eq:final-time-constraint}) into the objective
consists in introducing a Lagrange multiplier~${\cal F} \in \RR^{d'}$. This results in
the new objective
\begin{align}
L[p, {\cal F}] = S[p] + \left({\cal F}, O[u[p](\cdot, 0)] - a  \right)_{d'}.
\label{eq:lagrange-function}
\end{align}
Essentially, the additional
parameter~${\cal F}$ has to be tuned
in such a way that the final time constraint of $u$
is fulfilled. Note that a minimum of $S$ will always be at a saddle point
of $L$, since the constraint in~(\ref{eq:lagrange-function}) is added linearly.

In the context of the one-dimensional stochastic Burgers equation,
Chernykh and Stepanov (``CS'', \cite{chernykh-stepanov:2001}) used the method
of Lagrange multipliers to compute instantons for the PDF of the velocity
gradient $O[u(\cdot,0)] = \partial_x u(0,0)$.
They propose to
keep~${\cal F}$ fixed during the solution of each optimisation problem, instead
of adjusting~${\cal F}$ iteratively during the
computation. The latter would be necessary for the computation of the minimiser for a
prescribed value of~$a$ that is defined \textit{a priori}. 
The approach of keeping~${\cal F}$ fixed can be interpreted as a specific
variant of the method of Lagrange multipliers, when applied to the problem of
computing multiple instantons over a range of possible observable values $a$
for the PDF estimate. As a result of inverting
the mapping between ${\cal F}$ and $a$, only a single optimisation problem
needs to be solved for computing $a$ for a given ${\cal F}$, whereas typically a 
\textit{series} of unconstrained optimisation problems needs to be solved for
fixing $a$ a priori. We will refer to this procedure as the CS approach or
CS method, which is summarised in algorithm~\ref{algo:cs}.

This approach works well for computing PDFs
over a range of values of $a$. In that case, one can solve a single
unconstrained optimisation problem for each~${\cal F}$ in a specified range and
simply check \textit{a posteriori} (after the solutions of these
unconstrained problems have been found) to which observable value~$a$ the 
computed solution $(p_{\text{I}},u_{\text{I}})$ corresponds.
If this approach is feasible, it is numerically
cheaper than solving a series of unconstrained problems in order to find
the minimiser for given values of~$a$. In general, however, there are several
problems with this approach, as the map~${\cal F} \mapsto a$ can become
multivalued or diverge.

In particular, this is relevant whenever the PDF that
we want to estimate displays heavy tails in the sense that~$\rho_O$ decays slower
than~$\exp\left\{-c \lVert a \rVert \right\}$, or more generally if the map
$a \mapsto S_{\text{I}}(a)$ fails to be convex, as discussed
in~\cite{alqahtani-grafke:2021}; there, nonlinear transformations of the observable
that convexify the action are proposed as a solution to this problem.

Regarding the 3D NSE application, we applied
the CS approach of varying~${\cal F}$ and obtaining~$a$ a posteriori only
for the axisymmetric instantons. In this case we did not 
expect either heavy tails or non-unique
instantons based on our numerical experiments
and therefore made use of the faster solution times compared
to the approaches described below.

\begin{algorithm}
\caption{CS method for the constrained instanton optimisation
problem~(\ref{eq:min-constrained-supmat})}
\label{algo:cs}
\begin{algorithmic}
\lws
\REQUIRE{\\
Fixed Lagrange multiplier ${\cal F} \in \RR^{d'}$;\\
Initial control $p^{(0)}$ (e.g.\ $p^{(0)} \equiv 0$, or random
initialisation, or result from previously solved problem);\\
Error tolerance $\delta$.}
\ENSURE{\\
Control $p^{(*)}$;\\
Action $S[p^{(*)}]$;\\
Observable value $a^{(*)}$ for $p^{(*)}$\\}
\lws
    \STATE{Starting from $p^{(0)}$, approximately determine a minimum
    $p^{(*)}$ of $L[p, {\cal F}]$ with
    \begin{align*}
    \norm{\frac{\delta L}{\delta p}\left[p^{(*)}, {\cal F}\right]}_{
    L^2(\Omega\times[-T,0],\RR^d)}
    < \delta\,,
    \end{align*}
    where $L$ is the target functional~(\ref{eq:lagrange-function}),
    and possible algorithms are described in section~\ref{appendix:grad-lbfgs}.}
\STATE{Evaluate $S[p^{(*)}]$ and $a^{(*)} = O[u[p^{(*)}](\cdot, 0)]$.}
\end{algorithmic}
\end{algorithm}

\subsection{Penalty method}
\label{appendix:subsec-penalty}

As an alternative to the introduction of a Lagrange multiplier for the final
time constraint~(\ref{eq:final-time-constraint}), one can also use a penalty approach:
We introduce a penalty parameter~$\mu > 0$ and consider the objective
\begin{align}
R[p, \mu] = S[p] + \frac{\mu}{2} \norm{O[u[p](\cdot, 0)] - a}_{d'}^2\,.
\label{eq:penalty-functional}
\end{align}
It is intuitively clear and can be proven rigorously that,
for~$\mu \to \infty$, minima of~$R$ will fulfil
the constraint~(\ref{eq:final-time-constraint}) for the prescribed value of $a$.
For a finite value of~$\mu$, there is a trade-off between minimising
the quadratic action functional $S$ and minimising 
the penalty for deviations of $O[u[p](\cdot, 0)]$ from $a$, which is measured by the
second term in \eqref{eq:penalty-functional}. For small values of $\mu$ we
essentially ignore the additional constraint.
We note that the
functional~(\ref{eq:penalty-functional}) is bounded from below by 0, which is
in contrast to the situation in~(\ref{eq:lagrange-function}).

Numerically, it is not possible
to take the limit~$\mu \to \infty$. Instead, one solves a series of unconstrained
optimisation problems with penalty parameters~$(\mu^{(m)} )_{m = 0, \dots, M}$,
where~$\mu^{(m)}$ increases with~$m$. Using the solution of the previous
problem as the initial configuration for the next iteration and verifying the
convergence of~$O[u[p](\cdot, 0)]$ towards~$a$, approximate solutions
to~(\ref{eq:min-constrained-supmat}) can be obtained. 
The choice of step sizes for the
penalty parameters~$\mu^{(m)}$, as well as the maximum penalty~$\mu^{(M)}$ and
the termination criterion for the solution of each individual unconstrained
optimisation problem need to be considered carefully.

Different to the Lagrange approach, for this approach one cannot (easily) avoid
having to solve multiple optimisation problems by taking a short-cut.
The penalty method is summarised in algorithm~\ref{algo:penalty}.

\begin{algorithm}
\caption{Penalty method for the constrained instanton
optimisation problem~(\ref{eq:min-constrained-supmat})}
\label{algo:penalty}
\begin{algorithmic}
\lws
\REQUIRE{\\
Target observable value $a \in \RR^{d'}$;\\
Initial control $p^{(0)}$ (e.g.\ $p^{(0)} \equiv 0$, or random initialisation, or
result from previously solved problem);\\
Increasing sequence $\mu^{(0)}, \dots, \mu^{(M)}$ of penalty parameters;\\
Decreasing sequence $\delta^{(0)}, \dots \delta^{(M)}$ of error tolerances for each
unconstrained minimisation problem.}
\ENSURE{\\
Control $p^{(M)}$;\\
Action $S[p^{(M)}]$;\\
Observable value $a^{(M)}$ for $p^{(M)}$\\}
\lws
\FOR{$m = 0, 1, 2, \dots, M-1$}
    \STATE{Starting from $p^{(m)}$, approximately determine a
    minimum $p^{(*)}$ of $R[p, \mu^{(m)}]$ with
    \begin{align*}
    \norm{\frac{\delta R}{\delta p}\left[p^{(*)}, \mu^{(m)}\right]}_{
    L^2(\Omega\times[-T,0],\RR^d)}
    < \delta^{(m)}\,,
    \end{align*}
    \hspace{1em}where $R$ is the quadratic target
    functional~(\ref{eq:penalty-functional}), and minimisation
    algorithms are described in section~\ref{appendix:grad-lbfgs}.}
    \STATE{Set $p^{(m+1)} \leftarrow p^{(*)}$.}
\ENDFOR
\STATE{Evaluate $S[p^{(M)}]$ and $a^{(M)} = O[u[p^{(M)}](\cdot, 0)]$.}
\end{algorithmic}
\end{algorithm}

\subsection{Augmented Lagrangian method}
\label{appendix:subsec-augmented}

As a third possibility, both of the previous methods can be combined. This results in
the so-called augmented Lagrangian method, which is the method that we have
applied for the fully three-dimensional NSE instanton computations.
In this approach, one introduces both a Lagrange
multiplier~${\cal F} \in \RR^{d'}$ and a
penalty parameter~$\mu > 0$, which yields the functional
\begin{align}
L_A[p, {\cal F}, \mu] = S[p]  + \left({\cal F}, O[u[p](\cdot, 0)] - a  \right)_{d'}
+ \frac{\mu}{2} \norm{O[u[p](\cdot, 0)] - a}_{d'}^2\,,
\label{eq:augmented-lagrangian-supmat}
\end{align}
which is again bounded from below.
As for the penalty approach, it is necessary
to solve a series of optimisation problems for an increasing sequence of
penalty parameters~$(\mu^{(m)} )_{m = 0, \dots, M}$ in order to obtain an approximate
solution of the constrained problem~(\ref{eq:min-constrained-supmat}) with a
prescribed value of~$a$. However, it can be shown that,
by updating the Lagrange multiplier ${\cal F}^{(m)}$ according to
\begin{align}
{\cal F}^{(m + 1)} = {\cal F}^{(m)} + \mu^{(m)}
\left(O[u[p^{(m)}](\cdot, 0)] - a \right) \,,
\end{align}
where~$p^{(m)}$ denotes the solution of the~$m$-th unconstrained optimisation
problem, the convergence of~$O[u[p^{(m)}](\cdot, 0)]$ towards~$a$ can be
accelerated. As a result, it suffices to
consider smaller values of~$\mu^{(M)}$ for this method compared to
the pure penalty approach. This in turn avoids potential issues with ill-conditioning caused by large penalty parameters. The augmented Lagrangian method is
summarised in algorithm~\ref{algo:augmented}.

\begin{algorithm}
\caption{Augmented Lagrangian method for the constrained
instanton optimisation problem~(\ref{eq:min-constrained-supmat})}
\label{algo:augmented}
\begin{algorithmic}
\lws
\REQUIRE{\\
Target observable value $a \in \RR^{d'}$;\\
Initial control $p^{(0)}$ (e.g.\ $p^{(0)} \equiv 0$,
or random initialisation, or result from previously solved problem);\\
Initial guess for the Lagrange multiplier ${\cal F}^{(0)} \in \RR^{d'}$;\\
Increasing sequence $\mu^{(0)}, \dots, \mu^{(M)}$ of penalty parameters;\\
Decreasing sequence $\delta^{(0)}, \dots \delta^{(M)}$ of
error tolerances for each unconstrained minimisation problem.}
\ENSURE{\\
Control $p^{(M)}$;\\
Action $S[p^{(M)}]$;\\
Observable value $a^{(M)}$ for $p^{(M)}$\\}
\lws
\FOR{$m = 0, 1, 2, \dots, M-1$}
    \STATE{Starting from $p^{(m)}$, approximately determine
    a minimum $p^{(*)}$ of $L_{\text{A}}[p, {\cal F}^{(m)}, \mu^{(m)}]$ with
    \begin{align*}
    \norm{\frac{\delta L_{\text{A}}}{\delta p}
    \left[p^{(*)}, {\cal F}^{(m)}, \mu^{(m)}\right]
    }_{L^2(\Omega\times[-T,0],\RR^d)} < \delta^{(m)}
    \end{align*}
    \hspace{1em}where $L_{\text{A}}$ is the quadratic target
    functional~(\ref{eq:augmented-lagrangian-supmat}), and minimisation
    algorithms are described in section~\ref{appendix:grad-lbfgs}).}
    \STATE{Set $p^{(m+1)} \leftarrow p^{(*)}$.}
    \STATE{Set ${\cal F}^{(m + 1)} \leftarrow {\cal F}^{(m)}
    + \mu^{(m)} \left(O[u[p^{(m)}](\cdot, 0)] - a \right)$.}
\ENDFOR
\STATE{Evaluate $S[p^{(M)}]$ and $a^{(M)} = O[u[p^{(M)}](\cdot, 0)]$.}
\end{algorithmic}
\end{algorithm}

\section{Gradient-based minimisation: Gradient descent and L-BFGS method}
\label{appendix:grad-lbfgs}

In the previous section, the solution of the constrained optimisation
problem~(\ref{eq:min-constrained-supmat}) was reduced to the solution of unconstrained
optimisation problems involving different functionals, which can all be
considered as special cases of
\begin{align}
L_{\text{A}}[p, {\cal F}, \mu] = \frac{1}{2} \int_{-T}^0 \dd t \;
\left(p, \chi * p \right)_{L^2(\Omega,\RR^d)} + \left({\cal F}, O[u[p](\cdot, 0)] - a
\right)_{d'}  + \frac{\mu}{2} \norm{O[u[p](\cdot, 0)] - a}_{d'}^2\,.
\label{eq:augmented-lagrangian-supmat-general}
\end{align}
We still have the PDE-constraint \eqref{eq:pde-forward} but have removed the final time constraint \eqref{eq:final-time-constraint}.

In this section, we discuss gradient-based algorithms to minimise this
objective numerically with respect to~$p$ for fixed parameters~$\mu > 0$
and~${\cal F} \in \RR^{d'}$. We will first derive in
section~\ref{appendix:grad-lbfgs}~\ref{appendix:subsec-inst-eq} the first order necessary
conditions at minima of~(\ref{eq:augmented-lagrangian-supmat-general}), which are typically
referred to as instanton equations in this context.
In the literature, these
instanton equations are often solved iteratively by repeated forward and
backward integration. We connect this approach to gradient-based minimisation
of~(\ref{eq:augmented-lagrangian-supmat-general}) by conveniently expressing the gradient
$\delta L_{\text{A}} / \delta p$ using an adjoint state in
section~\ref{appendix:grad-lbfgs}~\ref{appendix:subsec-adjoint}. Afterwards, in
section~\ref{appendix:grad-lbfgs}~\ref{appendix:subsec-grad-desc}, we describe the
standard gradient descent algorithm for minimising \eqref{eq:augmented-lagrangian-supmat-general}.
In section~\ref{appendix:grad-lbfgs}~\ref{appendix:subsec-lbfgs}, we present
the L-BFGS scheme. This algorithm has significantly better
convergence properties than the standard gradient descent, especially for
ill-conditioned problems. Just like gradient descent though, only gradient
evaluations are needed and it is not necessary to compute second-order information
in form of a Hessian, which is typically very expensive.
We conclude by presenting implementational details concerning the memory
usage in section~\ref{appendix:grad-lbfgs}~\ref{appendix:redstor}.

\subsection{Instanton equations}
\label{appendix:subsec-inst-eq}

Here, we derive the instanton equations, i.e.~the first order necessary
optimality conditions for minima of~(\ref{eq:augmented-lagrangian-supmat-general}).
In order to easily formulate these conditions, we introduce another, field-valued
Lagrange multiplier $z$, the so called \textit{adjoint variable}. We focus
on the case $\mu = 0$, i.e., only consider the method of Lagrange multipliers from
section~\ref{appendix:unconstrained}~\ref{appendix:subsec-lagrange-mult}.
To formally derive the optimality conditions, we define
\begin{align}
  \begin{split}
\tilde{L}[u,p,z,{\cal F}] &= \frac{1}{2} \int_{-T}^0 \dd t
\left(p, \chi * p \right)_{L^2(\Omega,\RR^d)} 
 + \int_{_-T}^0 \dd t \left(z, \partial_t u + N(u)
- \chi * p \right)_{L^2(\Omega,\RR^d)} \\
& \quad  + \left(z(\cdot, -T), u(\cdot,-T) - u_0\right)_{L^2(\Omega,\RR^d)}  + \left({\cal F}, O[u(\cdot, 0)] - a  \right)_{d'}\,.
\end{split}
\end{align}
The first order necessary conditions then imply the following conditions to hold 
\begin{align}
\begin{cases}
\frac{\delta \tilde{L}}{\delta u} =  - \partial_t z + \nabla N(u)^\top z = 0\,,
\quad z(\cdot, 0)
= - \left[\frac{\delta O}{\delta u}(u(0))\right]^\top {\cal F}\,,\\[.1cm]
\frac{\delta \tilde{L}}{\delta p} = \chi * \left[p - z\right] = 0\,,\\[.1cm]
\frac{\delta \tilde{L}}{\delta z} = \partial_t u + N(u) - \chi * p = 0\,,\quad u(\cdot,-T) = u_0\,,\\[.1cm]
\nabla_{\cal F} \tilde{L} = O[u(\cdot, 0)] - a = 0\,.
\end{cases}
\end{align}
Here, the transpose is taken with respect to the scalar product of the space
$L^2(\Omega,\RR^d)$ . Within our definition of $\chi^{-1}$, the second condition
implies that for stationary points $(\uu, \pp, z_{\text{I}}, {\cal F}_{\text{I}})$
of the Lagrange function, the equality $\pp = z_{\text{I}}$ holds. This results in
the instanton equations
\begin{align}
  \begin{cases}
    \partial_t \uu + N(\uu) = \chi * \pp, & \uu(\cdot, -T) = u_0\,,\\
    \partial_t \pp - \nabla N(\uu)^\top \pp = 0, & \pp(0)
    = - \left[\frac{\delta O}{\delta u}(\uu(0))\right]^\top {\cal F}_\text{I}\,.
  \end{cases} \label{eq:instanton-cs}
\end{align}
with ${\cal F}_\text{I}$ being chosen such that final time constraint
$O[u(\cdot, 0)] - a = 0$ is satisfied.

In~\cite{chernykh-stepanov:2001}, CS solved these saddle-point equations
numerically by an iterative forward and backward solution
of~(\ref{eq:instanton-cs}). The connection to gradient-based minimisation
methods will be reviewed in the next two sections.

\subsection{Adjoint state method}
\label{appendix:subsec-adjoint}

We continue by explaining how the gradient $\delta L_{\text{A}} / \delta p$
can be computed using the adjoint state method, which involves the solution
of a forward and a backward PDE. While this could also be derived from the
approach of the previous section (see e.g.~\cite{plessix:2006}), we
demonstrate here directly that this is a valid way of expressing the gradient
$\delta L_{\text{A}} / \delta p$ at any $p$, and not only at critical points of
$L_{\text{A}}$. Expanding $L_{\text{A}}$ to first order
in $\delta p$, we have
\begin{align}\label{eq: derivation grad}
&L_{\text{A}}[p + \delta p, {\cal F}, \mu] =
L_{\text{A}}[p, {\cal F}, \mu] + \int_{-T}^0 \dd t \;
\left(\delta p, \chi * p\right)_{L^2(\Omega,\RR^d)} \nonumber\\
& + \int_\Omega \dd^d x' \int_{-T}^0 \dd t \int_\Omega \dd^d x \;
\bigg( {\cal F} + \mu \left(O[u[p](\cdot, 0)] - a \right),
\left. \frac{\delta O}{\delta u} \right|_{u(0)}(x')
\underbrace{\frac{\delta u(x', 0)}{\delta p(x,t)}}_{=: J(x',t'=0;x,t)}
\delta p(x,t) \bigg)_{d'}\,.
\end{align}
We now further simplify the second term. In order to be able to express it in a
convenient formulation, we \textit{define}
the adjoint field $z:\Omega \times [-T,0] \to \RR^d$ as the solution of
\begin{align}
\begin{cases}
\partial_t z - \nabla N(u)^\top z = 0\,,\\
z(0) = - \left[\frac{\delta O}{\delta u}(u[p](\cdot,0))\right]^\top \left\{ {\cal F}
+ \mu \left(O[u[p](\cdot, 0)] - a \right) \right\}\,.
\end{cases} \label{eq:adjoint-field-def}
\end{align}
We now use $z$ to rewrite the inner integral of the second term in \eqref{eq: derivation grad} by first computing the expression
\begin{align}
&\partial_ {t'} \int_\Omega \dd^d x' \left(z(x',t'),
J(x',t';x,t) \delta p(x,t) \right)_d\nonumber \\
&= \int_\Omega \dd^d x' \left(\partial_{t'} z(x',t'), J(x',t';x,t)
\delta p(x,t) \right)_d + \left(z(x',t'), \left[\frac{\delta}{\delta p(x,t)}
\partial_{t'}u(x',t') \right] \delta p(x,t) \right)_d \nonumber\\
&= \int_\Omega \dd^d x' \left(\nabla N^\top(u(\cdot,t'))(x') z(x',t'),
J(x',t';x,t) \delta p(x,t) \right)_d \nonumber\\
& \quad + \left(z(x',t'), \left[- \nabla N(u(\cdot,t'))(x') J(x',t';x,t)
+ \chi(x-x') \delta(t-t') \right] \delta p(x,t) \right)_d \nonumber\\
&= \left((\chi * z)(x,t'), \delta p(x,t) \right)_d \delta(t-t')\,.
\label{eq:adjoint-deriv}
\end{align}
Now, integrating~(\ref{eq:adjoint-deriv}) from~$t' = -T$ to~$t' = 0$,
we get (with~$J(x', -T; x,t) = 0$)
\begin{align}
\int_\Omega \dd^d x' \left(z(x',0), J(x',0;x,t)
\delta p(x,t) \right)_d = \left((\chi * z)(x,t), \delta p(x,t) \right)_d\,.
\end{align}
Finally, using the final condition in~(\ref{eq:adjoint-field-def}) and combining that with \eqref{eq: derivation grad}, we arrive at
\begin{align}
\frac{\delta L_{\text{A}}}{\delta p} = \chi * (p - z)\,.
\end{align}

We observe that for a single evaluation of the gradient of~$L_{\text{A}}$ with respect to $p$, we need to
solve two PDEs: First, given the current control~$p$, we need to integrate the
nonlinear equations~(\ref{eq:pde-forward}) forward in time in order to
determine~$u[p]$. Afterwards, we need to integrate the linearised
equations~(\ref{eq:adjoint-field-def}) backwards in time from~$t=0$ to~$t = -T$.
Here, both the starting condition for~$z$ at $t=0$ and
the term~$\nabla N(u)^\top$ will in general depend on~$u$.
We note that due to the negative sign in front of
the term~$\nabla N(u)^\top z$ in the adjoint equations,
integrating these equations backwards in time is actually the natural and
numerically stable choice.

\subsection{Minimisation algorithms -- Gradient descent}
\label{appendix:subsec-grad-desc}

Being able to compute the gradient of the target functional, we can now discuss
gradient-based minimisation algorithms. Generally, we consider iterative minimisation algorithms, where the
approximation to the optimal control $p$ is updated as
\[
p^{(k+1)} = p^{(k)} + \sigma^{(k)} s^{(k)},
\]
with $s^{(k)}$ being the search direction and $\sigma^{(k)} \in \RR_+\backslash \{0\}$ being the step length.
We first consider the simplest approach for computing the search direction $s$, which is to
choose the direction of steepest descent given by $- \delta L_{\text{A}} / \delta p$. We precondition
the search direction~$s$ via
\begin{align}\label{eq:precon chi s}
s = -\chi^{-1} * \frac{\delta L_{\text{A}}}{\delta p} = - (p-z)\,.
\end{align}
To compute the step length $\sigma$, we use
Armijo line search with backtracking.
The full scheme is summarised in
algorithm~\ref{algo:gradient-descent} and will be discussed in more detail in the following.

The preconditioner used in \eqref{eq:precon chi s}, i.e.\ the correlation matrix, corresponds to the
Hessian or second variation of the functional~$L_{\text{A}}$ in the
linear case (i.e., when both~$N$ and~$O$ are linear and~$\mu = 0$).
It significantly improves
the convergence of the scheme in numerical experiments with rapidly decaying $\chi$.
Indeed, with this preconditioner, the update steps for~$p^{(k)}$ read
\begin{align}
p^{(k+1)} = p^{(k)} + \sigma^{(k)} s^{(k)} =
\left(1 - \sigma^{(k)} \right) p^{(k)} + \sigma^{(k)} z^{(k)}\,,
\label{eq:cs-update}
\end{align}
which is precisely the weighted update scheme of~\cite{chernykh-stepanov:2001} for an iterative solution of~(\ref{eq:instanton-cs}),
which has been presented as a fixed-point like iteration there.

Different to \cite{chernykh-stepanov:2001}, we compute the step length using Armijo line search with backtracking.
It can be shown that gradient descent with Armijo line search converges to stationary points of minimisation problems under
suitable conditions. For our specific case, the \textit{Armijo} condition, also called \textit{sufficient decrease} condition, is given by
\begin{align}
L_{\text{A}}\left[ p^{(k+1)}, {\cal F}, \mu \right]
\leq L_{\text{A}} \left[ p^{(k)},{\cal F}, \mu \right]
+ c \sigma^{(k)} \left(g^{(k)}, s^{(k)}\right)_{L^2(\Omega\times[-T,0],\RR^d)}\,,
\label{eq:sufficient-decrease}
\end{align}
where $g^{(k)}$ is the gradient at $p^{(k)}$ and $c>0$ is the sufficient decrease constant. Using the
definition~(\ref{eq:augmented-lagrangian-supmat-general}) of the target functional,
this condition can be rewritten as
\begin{align}
&\sigma^{(k)} \left(p^{(k)}, \chi * s^{(k)}\right)_{L^2(\Omega\times[-T,0],\RR^d)}
+ \frac{\left(\sigma^{(k)}\right)^2}{2} \left(s^{(k)}, \chi * s^{(k)}\right)_{
L^2(\Omega\times[-T,0],\RR^d)} \nonumber\\
&\leq \left({\cal F},O[u[p^{(k)}](\cdot, 0)] - O[u[p^{(k+1)}](\cdot, 0)] \right)_{d'}
+ \frac{\mu}{2} \left\{\norm{O[u[p^{(k)}](\cdot, 0)] - a}_{d'}^2 \right. \nonumber\\
&\left. \quad - \norm{O[u[p^{(k+1)}](\cdot, 0)] - a}_{d'}^2 \right\} + c \sigma^{(k)}
\left(g^{(k)}, s^{(k)}\right)_{L^2(\Omega\times[-T,0],\RR^d)}\,.
\end{align}
We note that all~$L^2$ scalar products in this inequality only need to be evaluated once for each
complete line search as they stay the same for different values of~$\sigma^{(k)}$. Due to the appearance of~$O[u[p^{(k+1)}](\cdot, 0)]$, however,
it is necessary to integrate~(\ref{eq:pde-forward}) once for each step
size~$\sigma^{(k)}$ that is considered in the line search loop.

The approach that we described above and that we applied for minimising 
\eqref{eq:augmented-lagrangian-supmat-general} is called \textit{optimise then discretise}: we
first set up the optimality system (on a continuous level) and then discretise each equation.
The alternative approach is called \textit{discretise then optimise}. Here, one first
discretises the optimal control problem, in particular the state equation, and then derives optimality conditions for the discretised
problem, which can be a very tedious thing to do. Ideally, both approaches commute.
If this is not the case, one could end up
with inconsistent gradient discretisations for the approach of \textit{optimise then discretise}, which potentially result in the break down of the gradient scheme.
Typically, increasing the resolution in this case results in better gradient approximations and therefore cures (part of) the problem. It is generally better though to address the problem by choosing the discretisation of the state and the adjoint equations very carefully, ideally in such a way to ensure commutativity of the approaches.

For the kind of problems considered here, one needs to make sure that the 
time stepping schemes for the forward and
backward equations and the quadrature rules for the time integral that appears in the $L^2$ integrals
are a good fit to each other. As an additional safeguard  for inconsistencies in the gradient computation,
we implemented a break condition for the case
that the step size considered by the line search algorithm becomes too small.
If the algorithm stops early due to this criterion, then one has to check
individually whether the current accuracy is enough for the application at
hand, or if an increased resolution is necessary, which often leads to a better gradient approximation.
In our case, it helped to increase the resolution of the time stepping scheme.

Finally, we remark that the most natural stopping criterion for the gradient
descent algorithm is to require that~$||g^{(k)}||_{L^2(\Omega\times[-T,0],\RR^d)}$
is smaller than a given threshold tolerance.
We therefore use this stopping criterion instead of checking whether the relative changes of the observable~$a^{(k)}$
from one step to the next are below a threshold,
as suggested in~\cite{grafke-grauer-schaefer:2015}.

\begin{algorithm}
\caption{Gradient descent for the minimisation 
of~(\ref{eq:augmented-lagrangian-supmat-general})}
\label{algo:gradient-descent}
\begin{algorithmic}
\lws
\REQUIRE{\\
Target observable value $a \in \RR^{d'}$;\\
Penalty parameter $\mu > 0$;\\
Lagrange multiplier ${\cal F} \in \RR^{d'}$;\\
Initial control $p^{(0)}$ (e.g.\ $p^{(0)} \equiv 0$,
or random initialisation, or result from previously solved problem);\\
Error tolerance $\delta$;\\
Maximum step number $K$;\\
Initial step size $\sigma_{\text{init}} > 0$ (typically $\sigma_{\text{init}} = 1$);\\
Minimum step size $\sigma_{\text{min}} \ll \sigma_{\text{init}}$;\\
Backtracking fraction $\beta \in (0,1)$ (e.g.\ $\beta = 1/2$);\\
Sufficient decrease constant $c >0$ (typically $c \approx 10^{-2}$).
}
\ENSURE{\\
Control $p^{(*)}$ (approximate minimum of $L_{\text{A}}$);\\
Gradient norm $||\delta L_{\text{A}} / \delta p(p^{(*)})
||_{L^2(\Omega\times[-T,0],\RR^d)}$
at the approximate minimum;\\
Augmented Lagrangian $L_{\text{A}}[p^{(*)}, {\cal F}, \mu]$;\\
Observable value $a^{(*)}$ for $p^{(*)}$\\}
\lws
\FOR{$k = 0, 1, 2, \dots, K-1$}
    \STATE{Compute the gradient at $p^{(k)}$:}
    \STATE{Starting from~$u_0$, integrate the
    forward equation~(\ref{eq:pde-forward}) with $p^{(k)}$ as RHS and store the
    solution~$u^{(k)}$.}
    \STATE{Store the current observable value $a^{(k)} = O[u[p^{(k)}](\cdot, 0)]$.}
    \STATE{Integrate~(\ref{eq:adjoint-field-def})
    backwards to get~$z^{(k)}$, and compute the gradient
    \begin{align*}
    g^{(k)} \leftarrow \chi * (p^{(k)} - z^{(k)})\,.
    \end{align*}}
    \STATE{Compute and store $||g^{(k)}||_{L^2(\Omega\times[-T,0],\RR^d)}$.}
    \IF{$||g^{(k)}||_{L^2(\Omega\times[-T,0],\RR^d)} < \delta$}
    \STATE{\textbf{break}}
    \ENDIF
    \STATE{Fix the search direction
    \begin{align*}
    s^{(k)} \leftarrow - \chi^{-1} * g^{(k)}
    \end{align*}
    \quad and perform an Armijo line search to determine the step length $\sigma^{(k)}$:}
    \FOR{$\sigma^{(k)} = \sigma_{\text{init}}, \beta \sigma_{\text{init}},
    \beta^2 \sigma_{\text{init}}, \dots$}
    \STATE{Set $p^{(k+1)} \leftarrow p^{(k)} + \sigma^{(k)} s^{(k)}$.}
    \STATE{Evaluate $L_A[p^{(k+1)},{\cal F}, \mu]$ (needs
    another forward integration).}
    \IF{$L_{\text{A}}[p^{(k+1)},{\cal F}, \mu] \leq L_{\text{A}}[p^{(k)},{\cal F}, \mu]
    + c \sigma^{(k)} (g^{(k)}, s^{(k)})_{L^2(\Omega\times[-T,0],\RR^d)}$}
    \STATE{\textbf{break} current loop.}
    \ENDIF
    \IF{$\sigma^{(k)} < \sigma_{\text{min}}$}
    \STATE{\textbf{break} current and outer loop; report.}
    \ENDIF
    \ENDFOR
\ENDFOR
\end{algorithmic}
\end{algorithm}

\subsection{Minimisation algorithms -- L-BFGS method}
\label{appendix:subsec-lbfgs}

While the gradient descent scheme is simple to realise, globally convergent, and robust, its convergence can be very slow if the problem
is ill-conditioned. Using the true Hessian of $L_{\text{A}}$ as a
preconditioner would result in a significant increase in convergence speed. This would correspond to
using an exact Newton scheme, which comes with quadratic convergence instead of the
linear convergence that the gradient descent scheme
offers (with convergence rates that might be very close to 1 for ill-conditioned problems). Practically, however, it is infeasible to explicitly compute and store all the entries of the
discrete Hessian of $L_{\text{A}}$ for the (3+1)-dimensional PDE-constrained optimisation problem in case of the 3D NSE.
If we stick to solvers that only require matrix-vector products instead of knowledge of the full matrix,
this might be an approach to examine as evaluating the Hessian or its inverse in certain directions is achievable.
We plan to do so in future work.

For now, we use \textit{Quasi Newton} schemes. These schemes typically show superlinear convergence. Therefore, they converge significantly
faster than the pure gradient scheme. Different to Newton-like approaches, it is not necessary to supply second-order information.
Instead, the scheme itself extracts curvature information out of the past steps. A very well-known Quasi Newton method is the so called
BFGS scheme. Given that we have to deal with a very high-dimensional optimisation problem here, we use a variant with limited memory requirements, the
so called L-BFGS scheme.

For this
scheme, one has to store the last $m$ control updates (with $m$ typically being chosen between 3 and 20)
\begin{align}
\Delta p^{(k-1)} := p^{(k)} - p^{(k-1)}, \dots, 
\Delta p^{(k-m)} := p^{(k-m+1)} - p^{(k-m)}
\end{align}
as well as the last $m$ gradient updates
\begin{align}
\Delta g^{(k-1)} := g^{(k)} - g^{(k-1)}, \dots, 
\Delta g^{(k-m)} := g^{(k-m+1)} - g^{(k-m)}.
\end{align}
From this 
information, a low-rank approximation of the inverse of the 
Hessian $H^{(k)} = \delta^2 L_{\text{A}}[p^{(k)}, {\cal F}, \mu]$ 
is computed. This approximation is directly applied to the gradient 
descent direction $-g^{(k)}$ as a preconditioner, without storing it explicitly.
We also note that as a result there is no need to solve linear systems when applying the preconditioner as
the algorithm directly computes an approximation to the \textit{inverse} Hessian.

The L-BFGS algorithm is derived as a variant of the standard BFGS scheme, which we will briefly sketch in the following.
For the computation of the approximate inverse Hessian, we start
with an initial symmetric and positive 
definite estimate $B^{(0)}$ for the inverse Hessian $H^{(0)}$ at 
the initial position.  Standard choices are $B^{(0)} = \text{Id}$ or a scaled version of that. We use $B^{(0)} = \chi^{-1}$.
The standard BFGS scheme for approximating the inverse $B^{(k+1)}$, which is then used to compute the
search direction $s^{(k+1)} = - B^{(k+1)} g^{(k+1)}$ at the $(k+1)$-th step of our iterative method, is given by
\begin{align}
B^{(k+1)} &= B^{(k)} + \rho^{(k)} \left[ w^{(k)} \otimes \Delta p^{(k)} 
+ \Delta p^{(k)} \otimes w^{(k)} \right] \nonumber\\
& \quad - \left[\rho^{(k)}\right]^2 \left(w^{(k)}, \Delta g^{(k)} \right)_{
L^2(\Omega \times [-T,0], \RR^d)} \Delta p^{(k)} \otimes \Delta p^{(k)} \nonumber\\
&= \left[V^{(k)}\right]^\top B^{(k)} V^{(k)} + \rho^{(k)}
\Delta p^{(k)} \otimes \Delta p^{(k)}\,,
\label{eq:bfgs-full}
\end{align}
with
\begin{equation}
\rho^{(k)} = \left[ \left(\Delta g^{(k)}, \Delta p^{(k)} \right)_{
L^2(\Omega \times [-T,0], \RR^d)} \right]^{-1}
\end{equation}
as well as
\begin{equation}
w^{(k)} = \Delta p^{(k)} - B^{(k)} \Delta g^{(k)}
\end{equation}
and
\begin{equation}
V^{(k)} = \text{Id} - \rho^{(k)} \Delta g^{(k)} \otimes \Delta p^{(k)}\,.
\end{equation}

Theoretical results typically assume that Powell-Wolfe line search is used, which is a fairly expensive algorithm for computing the step length.
This guarantees that the
matrices $B^{(k)}$ remains positive definite at all steps. In practice, the much cheaper Armijo line search typically works very well.
As one no longer has a guarantee about the positive definiteness of $B^{(k)}$, it is recommended to
implement safety checks to ensure that $s^{(k+1)}$ is indeed a descent direction, such as
the angle condition given by
\begin{align}
\left(-g^{(k)}, s^{(k)} \right)_{L^2(\Omega \times [-T,0], \RR^d)} 
\geq \tilde{c} \norm{g^{(k)}}_{L^2(\Omega \times [-T,0], \RR^d)} 
\norm{s^{(k)}}_{L^2(\Omega \times [-T,0], \RR^d)},
\label{eq:angle-condition}
\end{align}
with a small constant $\tilde{c}$. This condition guarantees that the angle between the search direction
$s^{(k)}$ and the negative gradient $-g^{(k)}$ is smaller than 90 degrees.
If the angle condition is violated, a possible remedy would be to reset
$B^{(k)}$ to the initial estimate $B^{(0)}$, i.e.~to completely reset the BFGS memory.

Using the above formula for $B^{(k+1)}$ for large-scale problems is not recommended as the approximation matrix is usually dense.
We therefore rewrite the algorithm to deduce a limited memory version.
From the last line of~(\ref{eq:bfgs-full}), we see that the BFGS preconditioner can 
be written recursively as
\begin{align}
B^{(k+1)} &= \left[V^{(k)}\right]^\top B^{(k)} V^{(k)}
+ \rho^{(k)} \Delta p^{(k)} \otimes \Delta p^{(k)}\nonumber\\
&= \left[V^{(k-1)} V^{(k)}\right]^\top B^{(k-1)}
\left[V^{(k-1)} V^{(k)}\right] \nonumber\\
&\quad + \rho^{(k-1)} \left[V^{(k)}\right]^\top \Delta p^{(k-1)}
\otimes \Delta p^{(k-1)} V^{(k)} + \rho^{(k)} \Delta p^{(k)} \otimes \Delta p^{(k)}\\
&= \dots = \nonumber\\
&= \left[V^{(k-m+1)} \dots V^{(k)}\right]^\top B^{(k-m+1)}
\left[V^{(k-m+1)} \dots V^{(k)}\right]\nonumber\\
&\quad + \rho^{(k-m+1)} \left[V^{(k-m+2)} \dots V^{(k)}\right]^\top
\Delta p^{(k-m+1)} \otimes \Delta p^{(k-m+1)}
\left[V^{(k-m+2)} \dots V^{(k)}\right]\nonumber\\
&\quad + \rho^{(k-m+2)} \left[V^{(k-m+3)} \dots V^{(k)}\right]^\top
\Delta p^{(k-m+2)} \otimes \Delta p^{(k-m+2)}
\left[V^{(k-m+3)} \dots V^{(k)}\right]\nonumber\\
&\quad + \dots + \rho^{(k)} \Delta p^{(k)} \otimes \Delta p^{(k)}\,.
\label{eq:two-loop}
\end{align}
That means, we can write $B^{(k+1)}$ using the matrix $B^{(k-m+1)}$ and vector products that involve
$\Delta p^{(k)},\ldots, \Delta p^{(k-m+1)}$ as well as $\Delta g^{(k)}, \ldots, \Delta g^{(k-m+1)}$.

For the  L-BFGS scheme one now replaces the potentially dense matrix $B^{(k-m+1)}$ by an initial sparse guess
$B^{(k+1)}_0$, which can vary from iteration to iteration. Note that is not necessary to store full matrices.
Only the last $m$ entries for $\Delta p$ and $\Delta g$ need to be
stored in order to construct the preconditioner. A popular choice for the initial
guess would be
\begin{align}
B^{(k+1)}_0 = \gamma^{(k)} \; \text{Id} = \frac{
\left(\Delta p^{(k)}, \Delta g^{(k)} \right)_{L^2(\Omega \times [-T,0], \RR^d)}}
{\left(\Delta g^{(k)}, \Delta g^{(k)} \right)_{L^2(\Omega \times [-T,0], \RR^d)}}
\;\text{Id}\,.
\end{align}
Based on numerical experiments for the 1D viscid Burgers equation, 
we choose 
\begin{align}
B^{(k+1)}_0 = \gamma^{(k)} \; \chi^{-1}
\label{eq:lbfgs-init}
\end{align}
for the problem at hand. The linear operator $B^{(k+1)}$ can be applied to
the negative gradient $-g^{(k+1)}$ to compute the search direction $s^{(k+1)}$ without
explicitly storing it via the two-loop recursion that is summarised in
algorithm~\ref{algo:two-loop}. The algorithm for the full L-BFGS method is
shown in algorithm~\ref{algo:lbfgs}.

Formally, we lose the superlinear convergence property by using this limited memory version. Practically, we still observe dramatic speed-up compared to using a plain gradient descent. Compared to the standard BFGS scheme, we observe a reduction of memory requirements and operations counts by using the L-BFGS scheme.

\begin{algorithm}
\caption{Two-loop recursion for the evaluation of $s^{(k)} = -B^{(k)}g^{(k)}$}
\label{algo:two-loop}
\begin{algorithmic}
\lws
\REQUIRE{\\
Current gradient $g^{(k)}$;\\
$m$ previous gradient updates $\Delta g^{(k-1)}, \dots, \Delta g^{(k-m)}$;\\
$m$ previous control updates $\Delta p^{(k-1)}, \dots, \Delta p^{(k-m)}$;\\
$m$ previous weights $\rho^{(k-1)}, \dots, \rho^{(k-m)}$.
}
\ENSURE{\\
$s^{(k)} = -B^{(k)}g^{(k)}$ given by~(\ref{eq:two-loop}).\\}
\lws
\STATE{$r \leftarrow -s^{(k)}$.}
\FOR{$i = k-1, \dots, k-m$}
    \STATE{$\alpha^{(i)} \leftarrow \rho^{(i)} \left(\Delta p^{(i)},r 
    \right)_{L^2(\Omega \times [-T,0], \RR^d)}$.}
    \STATE{$r \leftarrow r - \alpha^{(i)} \Delta g^{(i)}$.}
\ENDFOR
\STATE{$r \leftarrow B^{(k)}_0 r$}
\FOR{$i = k-m, \dots, k-1$}
    \STATE{$h \leftarrow \rho^{(i)} \left(\Delta g^{(i)},r 
    \right)_{L^2(\Omega \times [-T,0], \RR^d)}$.}
    \STATE{$r \leftarrow r + \left[\alpha^{(i)} - h\right] \Delta p^{(i)}$.}
\ENDFOR
\STATE{return $r$.}
\end{algorithmic}
\end{algorithm}
\begin{algorithm}
\caption{L-BFGS method for the minimisation of~(\ref{eq:augmented-lagrangian-supmat-general})}
\label{algo:lbfgs}
\begin{algorithmic}
\lws
\REQUIRE{\\
Target observable value $a \in \RR^{d'}$;\\
Penalty parameter $\mu > 0$;\\
Lagrange multiplier ${\cal F} \in \RR^{d'}$;\\
Initial control $p^{(0)}$ (e.g.\ $p^{(0)} \equiv 0$,
or random initialisation, or result from previously solved problem);\\
Error tolerance $\delta$;\\
Maximum step number $K$;\\
Initial step size $\sigma_{\text{init}} > 0$ (typically $\sigma_{\text{init}} = 1$);\\
Minimum step size $\sigma_{\text{min}} \ll \sigma_{\text{init}}$;\\
Backtracking fraction $\beta \in (0,1)$ (e.g.\ $\beta = 1/2$);\\
Sufficient decrease constant $c >0$ (typically $c \approx 10^{-2}$);\\
Number of stored previous updates $m$ (typically $3$ to $20$);\\
Angle condition constant $\tilde{c}$ (typically $\tilde{c} \approx 10^{-4}$).
}
\ENSURE{\\
Control $p^{(*)}$ (approximate minimum of $L_{\text{A}}$);\\
Gradient norm $||\delta L_{\text{A}} / \delta p(p^{(*)})
||_{L^2(\Omega\times[-T,0],\RR^d)}$
at the approximate minimum;\\
Augmented Lagrangian $L_{\text{A}}[p^{(*)}, {\cal F}, \mu]$;\\
Observable value $a^{(*)}$ for $p^{(*)}$\\}
\lws
\FOR{$k = 0, 1, 2, \dots, K-1$}
    \STATE{Compute the gradient $g^{(k)}$ at $p^{(k)}$ as in
    algorithm~\ref{algo:gradient-descent}.}
    \IF{$k\geq m + 1$}
    \STATE{Delete $\Delta g^{(k-m-1)}$, $\Delta p^{(k-m-1)}$ and $\rho^{(k-m-1
    )}$}
    \ENDIF
    \IF{$k \geq 1$}
    \STATE{Store $\Delta p^{(k-1)}$, $\Delta g^{(k-1)}$ and $\rho^{(k-1)}
    \leftarrow 1 /  \left(\Delta g^{(k-1)}, \Delta p^{(k-1)} \right)_{
    L^2(\Omega \times [-T,0], \RR^d)}$.}
    \ENDIF
    \STATE{Compute and store $||g^{(k)}||_{L^2(\Omega\times[-T,0],\RR^d)}$.}
    \IF{$||g^{(k)}||_{L^2(\Omega\times[-T,0],\RR^d)} < \delta$}
    \STATE{\textbf{break}}
    \ENDIF
    \STATE{Fix the search direction with algorithm~\ref{algo:two-loop}:\;
    $s^{(k)} \leftarrow \text{Two-loop}\left( -g^{(k)}, 
    \Delta g, \Delta p, \rho \right)$, and~(\ref{eq:lbfgs-init}) as initial guess 
    for $B_0^{(k)}$ (use $s^{(k)} \leftarrow - \chi^{-1} * g^{(k)}$ for $k = 0$).}
    \IF{(\ref{eq:angle-condition}) is violated}
    \STATE{Reset to $s^{(k)} \leftarrow - \chi^{-1} * g^{(k)}$ 
    and delete all $\Delta g, \Delta p, \rho$.}
    \ENDIF
    \STATE{Perform an Armijo line search for $s^{(k)}$ to determine the step
    length $\sigma^{(k)}$ as in algorithm~\ref{algo:gradient-descent},
    and update $p^{(k+1)} \leftarrow p^{(k)} + \sigma^{(k)} s^{(k)}$.}
\ENDFOR
\end{algorithmic}
\end{algorithm}
\subsection{Memory reduction techniques}
\label{appendix:redstor}
Here, we briefly comment on memory reduction techniques that allow us to fit a
complete optimisation problem of the kind described above
in 3 spatial dimensions and 1 time dimension 
onto a single modern GPU without having to resort to
domain decomposition. These points have been previously discussed and quantitatively
analysed in detail in~\cite{grafke-grauer-schindel:2015}. Therefore, we only provide a few remarks in the following
about how to extend the ideas presented there to the slightly different setting that we use here.

First, as indicated in section~\ref{appendix:constrained}~\ref{appendix:subsec-oc},
for highly singular forcing correlation functions $\chi$, the effective
dimensionality of the control is rather small: Assuming that $\chi$ only acts on
large scales, its Fourier transform $\hat{\chi}(k)$ will only be non-zero for a
small and \textit{resolution-independent} number of modes $n_{k, \text{eff}}$.
In practice, one may either specifically choose a covariance that is strictly
set to 0 for all but a finite number of modes, or truncate a given $\chi$ at
modes where a suitable matrix norm of $\hat{\chi}(k)$ drops below a threshold.
The crucial observation is then that in the right-hand side of the state 
equation~\eqref{eq:pde-forward}, which is given by 
$\widehat{\chi * p}(k) = \hat{\chi}(k) \hat{p}(k)$, \textit{and} in the
action function
\begin{align}
S[p] = \frac{1}{2} \int_{-T}^0 \dd t \left(p, \chi * p \right)_{L^2(\Omega,\RR^d)}
=  \frac{1}{2} \int_{-T}^0 \dd t \; (2 \pi)^d\sum_{k \in \ZZ^d}
\left(\hat{p}(k), \hat{\chi}(k) \hat{p}(k) \right)_d,
\end{align}
where we used Parseval's identity, only these $n_{k, \text{eff}}$ modes enter
into the problem. Consequently, it is only necessary to store this reduced
number of modes. Note in particular that this approach is still valid for the
gradient computation and L-BFGS updates. Therefore, it suffices to store the
reduced mode representation for the gradient~$g^{(k)}$ in each time step, as well
as for the previous updates~$\Delta p^{(k)}$ and~$\Delta g^{(k)}$.

Even though the control only acts on a small number of Fourier modes,
the velocity field~$u$ and the adjoint field~$z$ will still be nonzero at
all appearing, small-scale modes, because the state equation is nonlinear
in general, and the final condition of the linear adjoint equation may also
involve small scales. Hence, even though only the large-scale~$n_{k, \text{eff}}$
modes of the gradient are actually used for the updates in the
optimisation algorithms, it is necessary to integrate the forward and backward
equations~(\ref{eq:pde-forward}) and~(\ref{eq:adjoint-field-def}) on the full
grid for each gradient computation. This causes the main costs of the algorithm
and is responsible for the scheme being expensive despite the discrete effective 
control space that we optimise over being comparatively low-dimensional.

Further, in order to be able to
integrate the adjoint equation~(\ref{eq:adjoint-field-def}), the full space-time history
of~$u[p^{(k)}]$ is needed. We avoid having to store this by using a checkpoint strategy for~$u$.
We only store  instances $u(\cdot, t_k)$ on a
logarithmically spaced subgrid in time during the forward integration. The price to pay is to do some redundant computations to
compute the information in between, which has not been stored, when it is needed for solving the adjoint equation.
Using this logarithmic scaling results in a good balance between saving memory and unnecessary recomputations.

Finally, we remark that the low dimension of the control space suggests that a
suitably modified implementation of Newton's method for the minimisation
of~(\ref{eq:augmented-lagrangian-supmat-general}) could be comparatively cheap.
In particular, the determinant of the reduced Hessian
of the action functional at the minimiser could be computed in this manner,
which would be relevant for numerically computing prefactor estimates to
improve~(\ref{eq:rho-log-asymp}). Extensions of the presented algorithms in
this direction are left as future work for now (also see the related
work~\cite{tong-vanden-eijnden-stadler:2020}).

\section{Three-dimensional Navier-Stokes instanton implementation}
\label{appendix:3d-nse-impl}

The SPDE with which we are concerned in the main text are the 3D NSE in nondimensionalised units
\begin{align}
\begin{cases}
    \partial_t u + (u\cdot\nabla)u = -\nabla P + \Delta u + \sqrt{\eps}\eta\,,\\
    \nabla\cdot u = 0\,,\\
    u(\cdot, -T) = u_0\,,
  \end{cases}
  \label{eq:nse-rescaled-supmat}
\end{align}
on $\Omega \times [-T,0]$ with $\Omega = [0, 2 \pi]^3$ with periodic boundary
conditions, a strictly solenoidal forcing with $\nabla \cdot \eta = 0$ for each
realisation, and initial condition $u_0 = 0$.

\subsection{Derivation of the adjoint equation}
\label{appendix:subsec-projection}

In order to directly apply the formalism that has
been developed in the previous sections to this example, it is necessary to remove
the additional incompressibility constraint $\nabla \cdot u = 0$ from the system.
In this section, we will show how this can be done by working with compressible
fields as a technical detour. The final optimisation algorithms will, however, be
formulated only in terms of incompressible control, state and adjoint variables.

In the incompressible NSE, the constraint $\nabla \cdot u = 0$ on the velocity
field is of course enforced through the pressure $P$, which can be eliminated by
taking the divergence of the evolution equation for $u$ and solving for $P$,
leading to
\begin{align}
P = - \Delta^{-1} \nabla \cdot ((u\cdot\nabla)u)\,.
\end{align}
Using the Leray projection $\PP$ onto the divergence-free part of a vector field
(see e.g.~\cite{majda-bertozzi:2001}),
which can be written as
\begin{align}
\PP = \text{Id} - \nabla \Delta^{-1} \nabla \cdot\,,
\end{align}
or
\begin{align}
\PP = \text{Id} - \frac{k k^\top}{\lVert k \rVert^2}
\end{align}
in Fourier space, the incompressible NSE can be reduced to
\begin{align}
\begin{cases}
    \partial_t u + \PP[(u\cdot\nabla)u] = \Delta u + \sqrt{\eps}\eta\,,\\
    u(\cdot, -T) = u_0\,.
  \end{cases}
  \label{eq:nse-proj1}
\end{align}
Here, for an initial condition that is divergence-free, the evolution equation
then obviously preserves solenoidality. Taking this approach one step further, we
may also consider the SPDE
\begin{align}
\begin{cases}
    \partial_t v + (\PP[v] \cdot \nabla)\PP[v] = \Delta v + \sqrt{\eps}\eta\,,\\
    v(\cdot, -T) = v_0
  \end{cases}
  \label{eq:nse-proj2}
\end{align}
for a compressible field $v$ without any restrictions on its divergence. In this
formulation, the projected field $u := \PP[v]$ will solve the incompressible NSE
with initial condition $u_0 = \PP[v_0]$. The advantage of~(\ref{eq:nse-proj2}) is
that it is straightforward to compute the transposed differential of the drift term
that is needed for the evolution equation for $z$ in the adjoint state method.
Similarly, in the field-theoretic formulation that has been briefly discussed in
section~\ref{appendix:constrained}~\ref{appendix:subsec-setup}, it is not necessary
to restrict the integration in path space to solenoidal vector fields through the
introduction of a functional Lagrange multiplier in this formulation. Now, with
\begin{align}
N(v) = (\PP[v] \cdot \nabla)\PP[v] - \Delta v\,,
\end{align}
and using the fact that $\PP$ is self-adjoint with respect to the
$L^2(\Omega, \RR^3)$ scalar product for periodic vector fields, we obtain
\begin{align}
&\left( \nabla N(v)^\top w, \delta v  \right)_{L^2(\Omega, \RR^3)}
= \left(w, \left.\frac{\mathrm{d}}{\mathrm{d} h}\right|_{h = 0}
N(v + h \delta v) \right)_{L^2(\Omega, \RR^3)} \nonumber\\
&= \left(w, (\PP[\delta v] \cdot \nabla) \PP[v] + (\PP[v] \cdot \nabla) \PP[\delta v]
- \Delta \delta v \right)_{L^2(\Omega, \RR^3)} \nonumber\\
&= \int_{\Omega} \dd^3x \; w_i \left(\PP[\delta v]_j \partial_j \PP[v]_i + \PP[v]_j
\partial_j \PP[\delta v]_i - \partial_j \partial_j \delta v_i \right)\nonumber \\
&= \int_{\Omega} \dd^3x \; \left[ - \PP[v]_i \partial_j w_i \PP[\delta v]_j
- \PP[v]_j \partial_j w_i \PP[\delta v]_i - \delta v_i \partial_j \partial_j w_i
\right]\nonumber \\
&= \left(-\PP \left[(\PP[v] \cdot \nabla) w  + (\nabla w)^\top \PP[v]\right]
- \Delta w, \delta v \right)_{L^2(\Omega, \RR^3)}\,.
\end{align}
Despite this excursion using compressible vector fields, the instanton
equations~(\ref{eq:instanton-cs}) and the optimisation algorithms of the 
previous section can still be written entirely in terms of incompressible vector
fields. Indeed, since for a solenoidal forcing, we must have
$\nabla \cdot \chi = 0$ for its correlation matrix function in the sense
that $\partial_i \chi_{ij} = 0$ and $\partial_j \chi_{ij} = 0$, the right hand
side of the state equation~(\ref{eq:pde-forward}) will automatically be solenoidal,
such that the PDE constraint on $u = \PP[v]$ can be written as
\begin{align}
\begin{cases}
\partial_t u + \PP[(u \cdot \nabla) u] - \Delta u = \chi * p\,,\\
u(\cdot, -T) = 0\,.
\end{cases}
\end{align}
Similar to the discussion in
section~\ref{appendix:grad-lbfgs}~\ref{appendix:redstor}, we can see that it
is only the solenoidal part of the control $p$ that enters the whole problem
and in particular the action functional $S[p] = \frac{1}{2} \int_{-T}^0 \dd t
\left(p, \chi * p \right)_{L^2(\Omega,\RR^3)}$, since the convolution with
$\chi$ will project $\hat{p}(k)$ onto $k^\top$ in Fourier space. As for the
adjoint equation, all observable quantities can only depend on $u = \PP[v]$,
such that the final condition for $z(\cdot, 0)$ can be written as
\begin{align}
z(\cdot, 0) = - \PP\bigg[ \left. 
\frac{\delta O}{\delta u}^\top \right|_{u(\cdot, 0)} \left\{ {\cal F}
+ \mu \left(O[u[p](\cdot, 0)] - a \right) \right\} \bigg]
\end{align} 
by the chain rule and is consequently solenoidal. The adjoint equation
\begin{align}
\partial_t z - \nabla N(u)^\top z = \partial_t z +\PP 
\left[(u \cdot \nabla) z  + (\nabla z)^\top u\right] + \Delta z = 0
\end{align}
will then preserve solenoidality. Summing up the discussion, the instanton
equations~(\ref{eq:instanton-cs}) at critical points of the action of the
incompressible NSE can be written as
\begin{align}
\begin{cases}
\partial_t u_{\mathrm{I}} + \PP \left[(u_{\mathrm{I}}
\cdot\nabla) u_{\mathrm{I}} \right] - \Delta u_{\mathrm{I}} = \chi * 
p_{\mathrm{I}}\,,\\
\partial_t p_{\mathrm{I}} + \PP \left[(u_{\mathrm{I}} \cdot\nabla) p_{\mathrm{I}} +
(\nabla p_{\mathrm{I}})^\top u_{\mathrm{I}}\right] + \Delta p_{\mathrm{I}} = 0\,,\\
p_{\mathrm{I}}(\cdot, t = 0) = - \PP\big[ \left. \frac{\delta O}{\delta u}^\top
\right|_{u_{\mathrm{I}}(0)} {\cal F}_{\mathrm{I}}\big]\,,
\end{cases}
\label{eq:inst-nse}
\end{align}
with $\nabla \cdot u_{\mathrm{I}} = \nabla \cdot p_{\mathrm{I}} = 0$.

\subsection{Forcing}
\label{appendix:subsec-forcing}

Let us now specify the forcing correlation that we chose and elaborate on
related implementation details. Imposing statistical homogeneity and isotropy
on the forcing, in addition to solenoidality, reduces the possible forms for
the correlation function to~\cite{robertson:1940}
\begin{align}
\chi(x) = f\left( \lVert x\rVert \right) \mathrm{Id} +
\frac{1}{2} \lVert x\rVert f'\left( \lVert x\rVert \right)
\left[\mathrm{Id} - \frac{x x^\top}{\lVert x\rVert^2} \right]\,,
\end{align}
where the only freedom is in the choice of the scalar function $f:[0, \infty]
\to \RR$. Corresponding to the physical picture of energy injection on large
scales, we work with the ``Mexican hat'' choice $f(r) = \chi_0 \exp \{-r^2 / 2
\lambda^2 \}$ with a forcing correlation length $\lambda > 0$ roughly the size
of the domain. In Cartesian coordinates, the matrix-valued function $\chi$ is
then given by
\begin{align}
\chi(x) = \frac{\chi_0}{2 \lambda^2} e^{-\frac{\lVert x\rVert^2}{2 \lambda^2}}
\left(
\begin{array}{ccc}
2 \lambda^2 - x_2^2 - x_3^2 & x_1 x_2 & x_1 x_3\\
x_1 x_2 & 2 \lambda^2 - x_1^2 - x_3^2 & x_2 x_3\\
x_1 x_3 & x_2 x_3 & 2 \lambda^2 - x_1^2 - x_2^2
\end{array} \right)
\label{eq:forcing-matrix}
\end{align}
for this choice of forcing correlation. For the pseudo-spectral NSE DNS and
instanton solver, we need the correlation in Fourier space, which is
\begin{align}
\hat{\chi}(k) = 2^{1/2} \pi^{3/2}\chi_0 \lambda^5
e^{-\frac{1}{2} \lambda^2 \lVert k\rVert^2} \left(
\begin{array}{ccc}
k_2^2 + k_3^2 & -k_1 k_2 & -k_1 k_3\\
-k_1 k_2 & k_1^2 + k_3^2 & -k_2 k_3\\
-k_1 k_3 & -k_2 k_3 & k_1^2 + k_2^2
\end{array} \right)\,.
\end{align}
We observe that the vector $k \in \RR^3$ is in the kernel of $\hat{\chi}(k)$,
so applying $\hat{\chi}$ will indeed render the output solenoidal. Furthermore,
on $k^\perp \subset \RR^3$, the matrix acts as the identity times
$c_k = 2^{1/2} \pi^{3/2}\chi_0 \lambda^5 \lVert k\rVert^2
e^{-\frac{1}{2} \lambda^2 \lVert k\rVert^2}$. In the reduced mode representation
of the control $p$, we concretely determine the number of
modes~$n_{k, \text{eff}}$ by keeping only those modes with $|c_k| >
\chi_{\text{tol}}$. As discussed in general in the previous sections, inverting
$\chi$ consists of discarding the possibly non-zero part of the input vector in
Fourier space parallel to $k$, and dividing the remaining input by $c_k$.

\begin{figure}
\centering
\includegraphics[width = .5\textwidth]{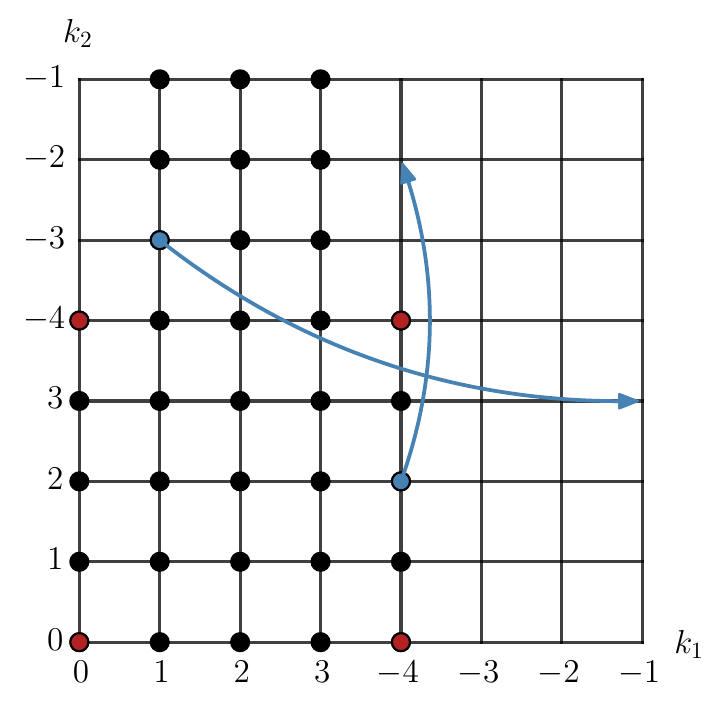}
\caption{Two-dimensional $8\times8$ grid in $k$-space. The 4 red points are
those where the Fourier transform of a real data set is real. These 4 points
plus the remaining 30 indicated points form a minimal subset that uniquely
determines the entire field, since each of the circled points $(k_1, k_2)$
possesses a partner at indices $(-k_1 , -k_2)$ that is determined by $\hat{\eta}
(-k_x , -k_y) = \hat{\eta}^*(k_x , k_y )$ . Two examples for this symmetry
are given by the blue points, with arrows pointing to their respective
partner. Note that one effectively mirrors each point $(k_1 , k_2)$ at any of
the red points to obtain the corresponding mode at $(-k_1 , -k_2)
\text{ mod } (8,8)$.}
\label{fig:kgrid}
\end{figure}

In the DNS of the stochastic NSE, we need to sample realisations of the
Gaussian random vector field $\eta$ with the prescribed correlation function
$\chi$. From the Fourier transform of the correlation in real space, we find
\begin{align}
\left<\hat{\eta}(k,t) \hat{\eta}^\dagger(k',t') \right>
= (2 \pi)^3 \hat{\chi}(k) \delta^{(3)}(k-k') \delta(t-t')\,,
\end{align}
i.e.~except for the trivial conjugation symmetry $\hat{\eta}(-k, t)
= \hat{\eta}(k,t)^*$ for real fields, the correlation in Fourier space
decouples between different modes for a stationary forcing. In order to
sample realisations of~$\hat{\eta}(k,t)$, it is, except for a
discretisation-dependent normalisation factor, hence enough to generate
independent and identically standard complex normally distributed random
vectors $Z(k,t) \sim {\cal C N}\left(0, \text{Id} \right)$ for all appearing
$k \in \ZZ^3$, multiply the result by a ``square root'' $\hat{\Lambda}(k)$ of
the forcing correlation $\hat{\chi}(k) = \hat{\Lambda}(k)
\hat{\Lambda}^\dagger(k)$, and enforce conjugation symmetry afterwards. The
random field $Z$ can be generated as $Z(k,t) = 2^{-1/2} (Z_r(k,t) + i Z_i(k,t))$
with independent $Z_r(k,t), Z_i(k,t) \sim {\cal N}\left(0, \text{Id} \right)$.
For our choice of $\chi$, the linear map $\hat{\Lambda}(k)$ simply corresponds
to projecting onto $k^\perp$ and multiplying by $\sqrt{c_k}$. All in all, if the
Fast Fourier Transform (FFT) implementation that is
used scales the inverse transformation by
$1/(n_{x_1} n_{x_2} n_{x_3})$, then the random forcing prior to the
symmetrisation step will be given by
\begin{align}
\hat{\eta}(k, t) = \left(\frac{n_{x_1} n_{x_2} n_{x_3}}{
2 \Delta t \Delta x_1 \Delta x_2 \Delta x_2} \right)^{1/2}
\Lambda(k) \left(Z_r(k,t) + i Z_i(k,t)\right)\,.
\end{align}
On a discrete grid in $k$-space with standard layout $k_i = 0, 1,
\dots, n_{x_i}/2-1, -n_{x_i}/2, \dots, -1$, the symmetry condition
$\hat{\eta}(-k, t) = \hat{\eta}(k,t)^*$ is understood modulo
$(n_{x_1}, n_{x_2}, n_{x_3})$, such that in addition to
$\hat{\eta}(k_1 = 0, k_2 = 0, k_3 = 0,t)$ being real, the same is also true
if any of the $k_j$ is equal to the Nyquist mode. As an example, the symmetry
constraints on a two-dimensional $8 \times 8$ grid are summarised in
figure~\ref{fig:kgrid}. On all grid points where the field is supposed to be
real, the a posteriori symmetrisation of the random field realisations
$\hat{\eta}$ has to set the imaginary part to zero and multiply the real part
by $\sqrt{2}$ in order to yield the desired covariance. For all other grid
points, one needs to identify symmetry partners at $k$ and $-k \text{ mod }
(n_{x_1}, n_{x_2}, n_{x_3})$, discard one of the two field values, say
$\eta(-k,t)$, and set it equal to $\eta(k,t)^*$. A more complete description
of sampling algorithms for Gaussian random fields using Fourier transforms can
be found in~\cite{lang-potthoff:2011}. Also note that if the vorticity
formulation of the incompressible NSE in terms of $\omega = \nabla \times u$
is used for the DNS, such that
\begin{align}
\partial_t \omega + \nabla \times (\omega \times u)
-\Delta \omega = \sqrt{\eps} (\nabla \times \eta)
\end{align}
and $\hat{u} = i k \times \hat{\omega} / \lVert k \rVert^2$, then the procedure
outlined above to sample $\hat{\eta}$ only needs to be modified by
multiplying $\hat{\eta}$ by an additional factor of $\lVert k \rVert$,
i.e.~$\widehat{\nabla \times \eta}(k,t) \overset{\text{law}}{=}
\lVert k \rVert \hat{\eta}(k,t)$.

\subsection{Observables and final conditions for the adjoint field}
\label{appendix:subsec-obs}

The two observables which we consider in the main text are both one-dimensional,
i.e.~$d'=1$ in the general setup of the previous sections, and correspond to the
longitudinal and transversal components of the velocity gradient tensor
$\nabla u$. Here, we derive the associated final time conditions for the adjoint
field. In order to do this, we go back to the formulation in terms of
compressible fields introduced in 
section~\ref{appendix:3d-nse-impl}~\ref{appendix:subsec-projection}.

For the first observable, the $z$-component of the vorticity 
\begin{align}
O_1[u(\cdot,0)] = (\nabla \times u)_z(x = 0, t = 0) = \omega_z(0,0)\,,
\end{align}
we rewrite this function as
\begin{align}
O_1[v(\cdot,0)] = (\nabla \times \PP[v])_z(x = 0, t = 0)
\end{align}
and compute, suppressing the time variable,
\begin{align}
&\left. \frac{\mathrm{d}}{\mathrm{d}h} \right|_{h = 0} O_1[v + h \delta v]
= \int_{\Omega} \dd^3 x \; \delta^{(3)}(x) (\nabla \times \PP[\delta v])_z
\nonumber\\
&= \int_{\Omega} \dd^3 x \; \left( - e_z \times \nabla \delta^{(3)}(x),
\PP[\delta v] \right) =  \left(- e_z\times \nabla \delta_0^{(3)},  \delta v\right)_{
L^3(\Omega, \RR^3)}\,,
\end{align}
where we used partial integration, the self-adjointness of $\PP$ as well as
the fact that $- e_z \times \nabla \delta_0^{(3)}$ is already solenoidal.
Hence, we obtain
\begin{align}
z_1(x, t = 0) = \left\{ {\cal F} + \mu \left(O[u[p](\cdot, 0)] - a \right)
\right\}  e_z \times \nabla \delta^{(3)}(x)\,
\end{align}
as the final condition for the adjoint field in the gradient
computation of~$L_{\text{A}}$.

For the second observable, the strain
\begin{align}
O_2[u(\cdot,0)] = \partial_z u_z(x = 0, t = 0)\,,
\end{align}
we proceed similarly and compute
\begin{align}
\left. \frac{\mathrm{d}}{\mathrm{d}h} \right|_{h = 0} O_2[v + h \delta v]
= \int_{\Omega} \dd^3 x \; \delta^{(3)}(x) \partial_z \PP[\delta v]_z 
= \left( - \PP\left[\partial_z \delta^{(3)}_0 e_z\right], \delta v \right)_{
L^3(\Omega, \RR^3)}\,,
\end{align}
such that
\begin{align}
z_2(\cdot, t = 0) = \left\{ {\cal F}
+ \mu \left(O[u[p](\cdot, 0)] - a \right) \right\} 
\PP\left[\partial_z \delta^{(3)}_0 e_z\right]
\end{align}
for the strain observable.

Both final conditions considered here are highly singular as they involve
derivatives of $\delta$ distributions, which is due to the fact that we
considered single-point observables without any averaging over a finite domain.
Of course, these conditions are easy to realise in a pseudo-spectral code in
Fourier space, where
\begin{align}
\hat{z}_1(k, t = 0) = i \left\{ {\cal F} + \mu \left(O[u[p](\cdot, 0)] - a \right)
\right\} \left(-k_y e_x + k_x e_y \right)
\end{align}
and
\begin{align}
\hat{z}_2(k, t = 0) = i \left\{ {\cal F} + \mu \left(O[u[p](\cdot, 0)] - a \right)
\right\} k_z \left( - \frac{k_x k_z}{\lVert k \rVert^2} e_x
-\frac{k_y k_z}{\lVert k \rVert^2} e_y
+ \left[1 - \frac{k_z^2}{\lVert k \rVert^2} \right] e_z  \right)\,.
\end{align}
We directly implemented these final conditions for the 3D code (with
additional $2/3$ anti-aliasing). However, these conditions are problematic
for different numerical schemes. In order to implement these conditions in
the finite difference code described below, it is necessary to instead consider
\begin{align}
\tilde{O}_1[u(\cdot,0)] = \left( K_{R} * \omega_z(\cdot, 0) \right)(0)
\label{eq:omz-smooth}
\end{align}
and
\begin{align}
\tilde{O}_2[u(\cdot,0)] = \left( K_{R} * \partial_z  u_z(\cdot, 0) \right)(0)
\label{eq:dzuz-smooth}
\end{align}
with a smooth kernel function $K_R$ in order to average over a ball $B_R(0)$
of radius $R > 0$ centred at the origin and consider the limit $R \to 0$ as
far as it is possible within the finite grid resolution. Details on the
resulting final conditions for $z$ in the axisymmetric finite difference code
are described below.

\subsection{Numerical implementation and parameters}
\label{appendix:subsec-3d-params}

The spatial discretisation of our full (3+1)-dimensional
optimisation code uses a $2/3$-dealiased pseudo-spectral method to evaluate
all nonlinearities and convolutions via FFTs. 
For the time discretisation, we
use the Heun scheme for both the forward and backward integration. The
diffusion term is integrated exactly using an integrating factor in Fourier
space in order to avoid the corresponding CFL restriction. All appearing time
integrals are discretised using the trapezoidal rule. As discussed in
section~\ref{appendix:grad-lbfgs}~\ref{appendix:subsec-grad-desc}, we followed
the \textit{optimise then discretise} approach, implying that with
these time discretisations, the adjoint method does not yield the exact
discrete gradient of the discrete action or augmented Lagrangian in our
specific example, and gradient-based descent algorithms may get stuck due to
this inconsistency. This inconsistency could be reduced in our case by increasing
the time resolution. As a result, the achieved accuracy turned out to be
sufficient for the PDF scaling estimates.

Since the instanton fields significantly grow in amplitude as $t \to 0$, we
use a non-uniform time parametrisation of the physical interval $[-T,0]$.
We keep this parametrisation fixed, such that the checkpointing algorithm
is easily implemented, and choose 
\begin{align}
\varphi:[-1,0] \to [-T,0], \quad \varphi(s) = t = -T 
\frac{e^{-t_\alpha s} - 1}{e^{t_\alpha} - 1}
\label{eq:time-parameterization}
\end{align}
with a constant $t_\alpha > 0$ and a uniform grid in $s$. While in principle 
we are interested in solving the minimisation
problem on the infinite time interval $T \to \infty$ in order to
characterise the stationary distribution of~$u$, we performed all
computations at finite~$T$ for this paper and
refrained from using a geometric 
reparametrisation~\cite{heymann-vanden-eijnden:2008, grafke-etal:2014}.
Instead, the time interval was chosen large enough such that it exceeds the
large eddy turnover time $T_{\text{LET}}$ at the Reynolds numbers which we
considered in DNS of the stochastic NSE.

All parameters of the full three-dimensional computations that have been
used to obtain the results shown in the main text are given in
table~\ref{tab:3d-params}. All necessary operators for the full 3D Navier-Stokes
solver used for this work have been implemented using the CUDA API and are
controlled through Python using PyCUDA~\cite{kloeckner-etal:2012}.

\begin{table}
\caption{Parameters of the full (3+1)-dimensional minimisation
of $L_{\text{A}}$, as defined in~(\ref{eq:augmented-lagrangian-supmat-general}),
for the incompressible 3D NSE~(\ref{eq:nse-rescaled-supmat}) with the L-BFGS method
as described in algorithm~(\ref{algo:lbfgs}).}
\label{tab:3d-params}
\centering
\begin{tabular}{L{3cm}L{5cm}R{3cm}}
\toprule
\textbf{Parameter} & \textbf{Explanation} & \textbf{Typical value}\\
\midrule 
$T$ & Time interval $[-T,0]$ & $1$\\
$\Omega$ & Periodic domain & $[0, 2 \pi]^3$\\
$u_0$   & Fixed initial condition for $u$ & $0$\\
$\chi_0$ & Default forcing strength in~(\ref{eq:forcing-matrix}) & $1$\\
$\lambda$ & Forcing correlation length in~(\ref{eq:forcing-matrix}) & $1$\\
$n_x$, $n_y$, $n_z$ & Spatial resolution & $128$\\
$n_t$ & Temporal resolution & $512$\\
$t_\alpha$ & Stretching factor for time in~(\ref{eq:time-parameterization})
& $3$\\
$\chi_{\text{tol}}$ & Threshold for reduced modes & $10^{-14}$\\
$\mu$ & Penalty parameters & $5, \dots, 400$\\
$m$ & Number of stored updates & $5$\\
$\delta$ & Error tolerance for gradient norm & $5 \cdot 10^{-4}$\\
$\sigma_{\text{init}}$ & Initial step size for line search & $1$\\
$\sigma_{\text{min}}$ & Minimum step size for line search & $5 \cdot 10^{-3}$\\
$\beta$ & Backtracking fraction & $0.5$\\
$c$ & Sufficient decrease constant in~(\ref{eq:sufficient-decrease}) & $10^{-2}$\\
$\tilde{c}$ & Angle condition constant in~(\ref{eq:angle-condition}) & $10^{-4}$\\
\bottomrule
\end{tabular}
\end{table}

\section{Axisymmetric Navier-Stokes instanton implementation}
\label{appendix:2d-axi-impl}

In this section, we include some implementation-related details of a second,
explicitly axisymmetric flow solver that has been used in order to compute
and stabilise axisymmetric solutions of~(\ref{eq:inst-nse}), as motivated in
the main text. In the following, we work in cylindrical coordinates
$(r,\theta,z)$ with $x_1 = r \cos \theta$, $x_2 = r \sin \theta$, $x_3 = z$.
For the axisymmetric instanton computations, we directly used the CS approach
as introduced in
section~\ref{appendix:unconstrained}~\ref{appendix:subsec-lagrange-mult},
i.e.~we only specify a range of Lagrange multipliers~${\cal F}$ without penalty,
and solve the instanton equations iteratively for fixed values of~${\cal F}$.

\subsection{Axisymmetric instanton equations and boundary conditions}
\label{appendix:subsec-axinst-bc}

We begin by noting that, while the forward and backward
evolution equations for $u_{\text{I}}$ and $p_{\text{I}}$ in~(\ref{eq:inst-nse})
are invariant under any rotation in $\RR^3$, the vorticity and strain
observables which we consider reduce this symmetry to axial symmetry under
rotation around the $z$-axis. It is hence natural to search for axisymmetric
solutions of~(\ref{eq:inst-nse}), which, however, are not guaranteed to yield
the global minimum of the corresponding optimisation problem. In order to make
the symmetry of the final conditions for $p_{\text{I}}$ more transparent,
we introduce, as indicated in
section~\ref{appendix:3d-nse-impl}~\ref{appendix:subsec-obs}, a smooth Gaussian
convolution kernel
\begin{align}
K_R(x) = \left(2 \pi R^2 \right)^{-3/2}
\exp\left\{- \frac{\lVert x \rVert^2}{2 R^2}\right\}
\end{align}
with radial extent $R > 0$ and consider the convolved
observables~(\ref{eq:omz-smooth}), (\ref{eq:dzuz-smooth}), which leads to the
final conditions
\begin{align}
p_{\text{I}}(r,\theta,z, t = 0) = -{\cal F}_{\text{I}}
\left(2 \pi R^2 \right)^{-3/2} R^{-2} r
\exp\left\{- \frac{r^2 + z^2}{2 R^2} \right\} e_\theta
\label{eq:omz-cylin}
\end{align}
for the $p$-field in the vorticity case and
\begin{align}
(\nabla \times p_{\text{I}})(r,\theta,z, t = 0) = -{\cal F}_{\text{I}}
\left(2 \pi R^2 \right)^{-3/2} R^{-4} r z
\exp\left\{- \frac{r^2 + z^2}{2 R^2} \right\} e_\theta
\label{eq:dzuz-cylin}
\end{align}
for the curl of the $p$-field in the strain case, where $e_\theta$ is the unit
vector in $\theta$-direction and axial symmetry is apparent. 

The axisymmetric instanton equations~(\ref{eq:inst-nse}) in cylindrical
coordinates are given by
\begin{align}
\begin{cases}
  D_t u_{r} - \frac{1}{r} u_{\theta}^2 + \partial_r P - L u_r = (\chi * p)_r\,,\\
  D_t u_{\theta} + \frac{1}{r}u_{r}u_{\theta} - L u_\theta = (\chi * p)_\theta\,,\\
  D_t u_{z} +\partial_z P - \left[\frac{1}{r}\partial_r(r\partial_r \cdot) 
  + \partial_{zz} \right]  u_z = (\chi * p)_z\,,\\[.3cm]
  D_t p_{r} - \frac{1}{r} u_{\theta} p_\theta + u_r \partial_r p_r 
  + u_\theta \partial_r p_\theta + u_z \partial_r p_z +\partial_r Q + L p_r = 0\,,\\
  D_t p_\theta + \frac{1}{r} (2 u_\theta p_r - u_r p_\theta) + L p_\theta = 0\,,\\
  D_t p_{z} + u_r \partial_z p_r + u_\theta \partial_z p_\theta 
  +  u_z \partial_z p_z + \partial_z Q + 
  \left[\frac{1}{r}\partial_r(r\partial_r \cdot) + \partial_{zz} \right] p_{z} = 0\,,
\end{cases}
\end{align}
where
\begin{align}
D_t = \partial_t + u_{r}\partial_r + u_{z}\partial_z
\end{align}
is the axisymmetric convective derivative. Here, $L$ denotes the elliptic operator
\begin{align}
L = \frac{1}{r}\partial_r(r\partial_r \cdot ) + \partial_{zz} - \frac{1}{r^2}
\end{align} 
originating from the vector Laplacian in cylindrical coordinates, and the fields
have to satisfy the incompressibility constraint
\begin{align}
\partial_ru_{r} + \frac{1}{r}u_{r} + \partial_z u_{z} = 0, \quad \partial_r p_{r}
+ \frac{1}{r} p_{r} + \partial_z p_{z} = 0\,,
\end{align}
which is enforced by the pressures $P$ and $Q$. In order to conveniently
incorporate the incompressibility constraint, we use streamfunctions with
\begin{align}
u = u(r,z) = \nabla \times (\psi(r,z) e_\theta) + u_\theta(r,z) e_\theta,
\quad p = p(r,z) = \nabla \times (\phi(r,z) e_\theta) + p_\theta(r,z) e_\theta\,.
\end{align}
The streamfunctions $\psi$ and $\phi$ determine the $r$ and $z$ components
of the fields via
\begin{align}
u_r = -\partial_z \psi, \quad u_z = \frac{1}{r} \partial_r (r \psi), \quad
p_r = - \partial_z \phi, \quad p_z = \frac{1}{r} \partial_r (r \phi)\,,
\end{align}
and can be computed from the vorticities $\omega = \nabla \times u$
and $\sigma = \nabla \times p$ by solving
\begin{align}
\omega_\theta = -L \psi, \quad \sigma_\theta = - L \phi\,.
\label{eq:l-stream}
\end{align}
Hence, taking $u_\theta$ and $\omega_\theta$ as well as $p_\theta$ and
$\rho_\theta$ as the independent dynamical variables, the instanton
equations in vorticity-streamfunction formulation are
\begin{align}
\begin{cases}
D_t u_\theta + \frac{1}{r} u_r u_\theta - L u_\theta = [\chi * p]_\theta\,,\\
D_t \omega_\theta - \frac{1}{r} u_r \omega_\theta - \frac{1}{r} \partial_z 
(u_\theta^2) - L \omega_\theta = [(\nabla \times \chi) * p]_\theta\,,\\
D_t p_\theta + \frac{1}{r} \left(2 u_\theta p_r - u_r p_\theta \right)
+ L p_\theta = 0\,,\\
D_t \sigma_\theta - \frac{1}{r} \partial_z (u_\theta p_\theta)+ \partial_z
u_\theta \partial_r p_\theta - \partial_r u_\theta \partial_z p_\theta \\
\quad +2 (\omega_\theta + 2 \partial_r u_z) \partial_r p_r + \frac{2}{r}
\partial_r u_z p_r + \left(2 \partial_z u_z + \frac{u_r}{r} \right)
\left(2 \partial_z p_r - \sigma_\theta \right) + L \sigma_\theta = 0\,,
\label{eq:inst-cylin-supmat}
\end{cases}
\end{align}
where the $r$ and $z$ components of $u$ and $p$ have to be computed by
solving~(\ref{eq:l-stream}). The final conditions for $p$
are~(\ref{eq:omz-cylin}) and~(\ref{eq:dzuz-cylin}) for the averaged observables
that we consider in the following, and in particular, we have
$\sigma_\theta(t=0) = 0$ for the vorticity observable and $p_\theta(t=0) = 0$ for
the strain observable. The instanton equations~(\ref{eq:inst-cylin-supmat}), to be
solved on the time interval $[-T,0]$ in the $r$-$z$-plane $[0,r_{\text{max}}] \times
[-z_{\text{max}},z_{\text{max}}]$, have to be supplemented 
by the following spatial boundary conditions: 
At the radial boundaries $r = 0$ and $r = r_{\text{max}}$, all fields obey
homogeneous Dirichlet boundary conditions (except for the $z$ components of the appearing vector fields at $r = 0$, which do not necessarily vanish there). In $z$, the boundary conditions are periodic
in $[-z_{\text{max}},z_{\text{max}}]$. 
We impose additional reflection symmetries at $z=0$ in
accordance with the equations of motion, which enables us to
solve~(\ref{eq:inst-cylin-supmat}) only in the upper half of the domain with $z \in
[0, z_{\text{max}}]$. 
Even fields with respect to reflection at $z = 0$ thus have to fulfil
homogeneous Neumann boundary conditions at $z = 0$ and $z = z_{\text{max}}$, whereas odd fields
with respect to reflection at $z = 0$ obey homogeneous Dirichlet boundary conditions in
$z$. Even symmetry can be imposed for $u_r$, $u_\theta$, $\omega_z$, $p_r$,
$p_\theta$ and $\sigma_z$, and odd symmetry for $u_z$, $\omega_r$, $\omega_\theta$,
$\psi$, $p_z$, $\sigma_r$, $\sigma_\theta$ and $\phi$.

\subsection{Spatio-temporal discretisation}

Based on the numerical approach of~\cite{grauer-sideris:1991} for the simulation
of axisymmetric Euler flows, we use standard symmetric second order finite
differences for all appearing spatial derivatives and discretise the spatial
domain on a regularly spaced grid of size $(n_r + 1) \times (n_z + 1)$ in
$[0,r_{\text{max}}] \times [0, z_{\text{max}}]$ with grid spacing $\Delta r =
r_{\text{max}} / n_r$ and $\Delta z = z_{\text{max}} / n_z$. 

For this spatial discretisation, the $r$ and $z$ components of the $u$ and $p$
fields at each instant in time are determined by first solving
\begin{align}
&-(L\psi)^{(n)}_{ij} =  -\left(\frac{\psi_{i+1,j}^{(n)} - 2 \psi_{i,j}^{(n)} +
\psi_{i-1,j}^{(n)}}{\Delta r^2}  + \frac{1}{r_{i}}\,  \frac{\psi_{i+1,j}^{(n)} -
\psi_{i-1,j}^{(n)}}{2 \Delta r} \right.\nonumber\\
&\hspace{3cm}\left. -\frac{1}{r_i^2} \, \psi_{i,j}^{(n)} +
\frac{\psi_{i,j+1}^{(n)} - 2 \psi_{i,j}^{(n)} +
\psi_{i,j-1}^{(n)}}{\Delta z^2}\right) = \omega_{\theta, ij}^{(n)}\,,
\label{eq:l-stream-discrete}
\end{align}
for $i = 1, \dots, n_r - 1$ and $j = 1, \dots, n_z - 1$ with Dirichlet 0
boundary conditions at $i = 0$, $i = n_r$ or $j = 0$ and $j = n_z$ for
$\psi_{ij}^{(n)}$. Here the upper index indicates fields at time $t_n$ and
lower indices denote the spatial position $r_i = i \Delta r$, $z_j = j \Delta z$.
Once we have determined $\psi^{(n)}$, which is discussed below, we can
determine $u_{r}^{(n)}$ and $u_{z}^{(n)}$ as
\begin{align}
u_{r,ij}^{(n)} = - \frac{\psi_{i,j+1}^{(n)} - \psi_{i,j-1}^{(n)}}{2 \Delta z}
\; , \;
u_{z,ij}^{(n)} = \frac{\psi_{i+1,j}^{(n)} - \psi_{i-1,j}^{(n)}}{2 \Delta r}
+ \frac{1}{r_i} \psi_{ij}^{(n)}\,,
\end{align}
where, for $j = 0$ and $j = n_z$, $\psi^{(n)}_{i,j+1}$ and $\psi^{(n)}_{i,j-1}$
are reflected to fulfil the imposed odd symmetry in $z$. For the time
discretisation of all evolution equations in~(\ref{eq:inst-cylin-supmat}), we use
the Leapfrog scheme. Suppose for now for notational simplicity that the time
interval $[-T,0]$ is discretised equidistantly with $n_t + 1$ points at distance
$\Delta t = T / n_t$. In the actual instanton computations, we again use the
exponentially stretched time parametrisation~(\ref{eq:time-parameterization}),
but it is straightforward to formulate the following discretisations for a
uniform grid in $s \in [-1, 0]$ and fields $\tilde{u} = u \circ \varphi$,
$\tilde{\omega} = \omega \circ \varphi$ and so forth. In order to avoid the
severe restriction on the permitted $\Delta t$ that would otherwise arise from
the diffusion CFL condition, we choose a Crank-Nicolson like, semi-implicit
discretisation $[L h^{(n+1)} + L h^{(n-1)}]/2$ for all four diffusion terms
in~(\ref{eq:inst-cylin-supmat}). Concretely, in order to determine $u_{\theta}^{(n+1)}$
given all fields at times $t^{(n-1)}$ and $t^{(n)}$, we discretise
\begin{align}
&\frac{u_{\theta,ij}^{(n+1)} - u_{\theta,ij}^{(n-1)}}{2 \Delta t} + u_{r,ij}^{(n)}
\frac{u_{\theta,i+1,j}^{(n)} - u_{\theta,i-1,j}^{(n)}}{2 \Delta r} + u_{z,ij}^{(n)}
\frac{u_{\theta,i,j+1}^{(n)} - u_{\theta,i,j-1}^{(n)}}{2 \Delta z} \nonumber \\
&+ \frac{1}{r_i} u_{r,ij}^{(n)} \frac{u_{\theta,i+1,j}^{(n)}
+ u_{\theta,i-1,j}^{(n)}}{2}  - \frac{1}{2} \left[
\left(Lu_{\theta}^{(n+1)} \right)_{ij}
+ \left(Lu_{\theta}^{(n-1)}\right)_{ij} \right]
= [\chi * p]_{\theta,ij}^{(n)}\,, \label{eq:udsicr}
\end{align}
such that a Helmholtz equation
\begin{align}
u_{\theta, ij}^{(n + 1)} - \Delta t  \left(Lu^{(n+1)}_\theta\right)_{ij} = r_{ij}
\label{eq:diff-implicit}
\end{align}
with an appropriate $r_{ij}$ as determined by~(\ref{eq:udsicr}) needs to be
solved to update $u_\theta$. All other equations in~(\ref{eq:inst-cylin-supmat})
are treated similarly. 

Both~(\ref{eq:l-stream-discrete}) and~(\ref{eq:diff-implicit}) are special
cases of a discretisation of the (noncritical) Helmholtz equation
\begin{align}
\begin{cases}
\left[a \; \text{Id} - b L\right] u = f, \quad L = \left[\frac{1}{r}
\partial_r(r \partial_r \cdot) - \frac{1}{r^2} + \partial_{zz} \right]
\text{ in } [0,r_{\text{max}}] \times [0, z_{\text{max}}]\,,\\
u(r = 0, \cdot) = u(r = r_{\text{max}}) = 0\,,\\
u(\cdot, z = 0) = u(\cdot, z = z_{\text{max}}) = 0 \text{ \textbf{or} }
\partial_z u(\cdot, z = 0) = \partial_z u(\cdot, z = z_{\text{max}}) = 0\,,
\end{cases}
\end{align}
with constants $a,b>0$. We use a multigrid solver for this equation with a
Gauss-Seidel smoother with one iteration per grid, full-weighting restrictions,
bilinear interpolation for the prolongation and W-cycles for the recursion
(see e.g.~\cite{trottenberg-etal:2000}), in order to solve the discretised
Helmholtz equations to an accuracy of $\epsilon_{\text{mgrid}} = 10^{-12}$.

Von Neumann stability analysis of the Leapfrog scheme for the linear advection
equation shows the existence of the so-called computational mode, leading
to even-odd-oscillations in the time domain at a frequency of $2/\Delta t$
of the scheme due to decoupling of directly adjacent fields in time. In order
to mitigate this problem, we added a Robert-Asselin-Williams time filter with
strength $10^{-2}$ to the scheme~\cite{robert:1966,asselin:1972,williams:2009}.

\subsection{Fast evaluation of polar convolutions}

Here, we discuss how to efficiently evaluate the convolutions $(\chi * p)_\theta$
and $(\chi * \sigma)_\theta$ which appear on the RHS of the forward equation
for $u$, as well as $(\chi * p)_r$ and $(\chi * p)_z$, which are additionally
needed for the computation of the action integral (which we discretise in
real space using the trapezoidal rule in space and time). 
In order to speed up the naive
${\cal O}\left(n^6 \right)$ scaling of a direct evaluation of a three-dimensional
convolution on a grid with $n$ points in each direction, the usual approach for
suitable domains is to employ the convolution theorem
\begin{align}
(f * g)(x) = \int_{\RR^3} \mathrm{d}^3 k \; \hat{f}(k) \hat{g}(k) e^{i (k,x)_3}
\end{align}
for $\Omega = \RR^3$ and $f,g:\Omega \to \RR$ integrable, or its discrete
equivalent using Fourier series on bounded, periodic domains. By virtue of the
FFT, this reduces the cost to compute a three-dimensional convolution to
${\cal O}\left( n^3 \log^3n \right)$. Since we consider axisymmetric fields as
well as an isotropic forcing and the convolved fields will thus be axisymmetric
as well, a valid option to compute the necessary convolutions in our setting
would be to first extrapolate all polar fields onto a full three-dimensional
grid. Afterwards, we could compute the convolutions with $\chi$ via 3D FFTs,
and then obtain the components of the result in cylindrical coordinates in the
$r$-$z$-plane e.g.~in the $x \geq 0$, $y = 0$ half-plane where $(\chi * p)_r =
(\chi*p)_x$ and $(\chi * p)_ \theta = (\chi*p)_y$. It is, however, possible to
directly evaluate the convolutions in cylindrical coordinates, which is in
particular cheaper for a large-scale forcing. The associated costs scale as
${\cal O}\left(n_{\text{Bz}} n^2 \log n  \right)$, where $n_{\text{Bz}}$ is the (small)
number of zeros of various Bessel functions appearing in the polar convolutions,
as detailed below. For a comprehensive introduction to the evaluation of Fourier
transforms and convolutions in polar coordinates, see~\cite{baddour:2009}.

The Fourier transform of a scalar function $f$ in cylindrical coordinates $(r, \theta, z)$
can in general be expressed as
\begin{align}
\hat{f}(k_r, \theta_k, k_z) &= \int_{0}^\infty \dd r \int_0^{2 \pi} 
\dd \theta \int_{-\infty}^\infty \dd z \; r f(r,\theta,z) e^{-ik_z z}
e^{-i k_r r \cos(\theta_k - \theta)} \nonumber\\
&= \sum_{l = - \infty}^\infty 2 \pi i^{-l} \HH_l\left[\tilde{f}_l
\right](k_r, k_z) e^{i l \theta_k} =: \sum_{l = - \infty}^\infty
\hat{f}_l(k_r, k_z) e^{i l \theta_k}\,, \label{eq:fourier-polar-general}
\end{align}
where
\begin{align}
\tilde{f}(r,\theta,k_z) = \int_{-\infty}^\infty \dd z \; f(r,\theta,z)
e^{-ik_z z} = \sum_{m =  -\infty}^\infty \tilde{f}_{m}(r, k_z) e^{i m \theta}
\end{align}
is the partial Fourier transform in $z$ direction, which we then expanded
in an angular Fourier series with coefficients
\begin{align}
\tilde{f}_m(r, k_z) = \frac{1}{2 \pi} \int_{0}^{2 \pi} \dd \theta\;
\tilde{f}(r, \theta, k_z) e^{-im\theta}\,.
\end{align}
Finally, inserting the expansion
\begin{align}
e^{-i k_r r \cos(\theta_k - \theta)} = \sum_{l = - \infty}^\infty
i^{-l} J_l\left( k_r r \right) e^{-il\theta} e^{il \theta_k}
\end{align}
in terms of Bessel functions $J_l$  of the first kind of order $l$ resulted
in the second line of~(\ref{eq:fourier-polar-general}), which involves $l$-th
order Hankel transforms, defined via
\begin{align}
\HH_l\left[h \right](k_r) := \int_0^\infty \dd r \; r h(r) J_l(k_r r)
\label{eq:def-hankel}
\end{align}
for functions $h: [0, \infty) \to \RR$ for which the integral exists. Also note
that for negative integer orders, the Bessel functions of the first kind fulfil
$J_{-l} = (-1)^l J_l$, $l \in \NN$. In summary, computing the Fourier transform
of a function in cylindrical coordinates consists of three steps: First, perform
a Fourier transform in $z$, then a Fourier series decomposition in $\theta$,
followed by radial Hankel transforms of order $l$ for each term of the angular
series. In the special case that $f$ is axisymmetric, the Fourier
transform~(\ref{eq:fourier-polar-general}) reduces to
\begin{align}
\hat{f}(k_r, k_z) = 2 \pi \HH_0\left[\tilde{f} \right](k_r, k_z)\,.
\end{align}
Since the Hankel transform is an involution for all real $l \geq -1/2$ and in
particular for all $l \in \ZZ$, inverse Fourier transforms in cylindrical
coordinates can analogously be evaluated by performing an ordinary inverse
transform in $z$ and Hankel transforms of order $l$ times $i^l/2\pi$ on all
angular coefficients, such that
\begin{align}
f(r,\theta,z) = \sum_{l = -\infty}^\infty f_l(r,z) e^{il\theta} =
\sum_{l = -\infty}^\infty e^{il\theta} \frac{i^l}{2 \pi} \HH_l
\left[\text{IFT}_z\left[\hat{f}_l \right]\right](r,z)\,.
\label{eq:ift-polar}
\end{align}

Now, we elaborate on how to use the polar Fourier transform that has just
been introduced for the convolution of $\chi$ and $p$ in practice. Let us focus
on the example of evaluating $(\chi * p)_\theta$ for concreteness. In order to
avoid integrations over unit vectors in cylindrical coordinates, we compute
$(\chi * p)_\theta(r,z)$ as
\begin{align}
(\chi *p)_\theta(r,z) = \left[\chi_{yx}*p_x + \chi_{yy}*p_y +
\chi_{yz} * p_z\right](x=r,y=0,z)
\end{align} 
using Cartesian components. For the three individual scalar convolutions
that need to be evaluated, we first analytically compute the Fourier transforms
of the components of $\chi$ (if possible) and expand them in an angular Fourier
series. For the specific choice~(\ref{eq:forcing-matrix}) for the forcing
covariance, we obtain e.g.
\begin{align}
\hat{\chi}_{yx}(k_r,\theta_k,k_z)&=-2^{1/2}\pi^{3/2}\chi_0 \lambda^5
e^{-\frac{1}{2}\lambda^2(k_r^2+k_z^2)} k_r^2 \sin \theta_k \cos \theta_k \nonumber\\
&= i 2^{-3/2} \pi^{3/2}\chi_0 \lambda^5 e^{-\frac{1}{2}\lambda^2(k_r^2+k_z^2)}
k_r^2 \left(e^{2 i \theta_k} - e^{-2i \theta_k} \right) \nonumber\\
&=\hat{\chi}_{yx,2} e^{2i \theta_k}+\hat{\chi}_{yx,-2} e^{-2i \theta_k}\,,\\
\hat{\chi}_{yy}(k_r,\theta_k,k_z)&= 2^{1/2} \pi^{3/2}\chi_0 \lambda^5
e^{-\frac{1}{2}\lambda^2(k_r^2+k_z^2)} (k_r^2 \cos^2 \theta_k + k_z^2) \nonumber\\
&= 2^{-3/2} \pi^{3/2} \chi_0 \lambda^5 e^{-\frac{1}{2}\lambda^2(k_r^2+k_z^2)}
\left(k_r^2 e^{2i \theta_k} + k_r^2 e^{-2i \theta_k} +(2 k_r^2 + 4 k_z^2)
e^{0 i \theta_k}\right)\nonumber\\
&=\hat{\chi}_{yy,2} e^{2i \theta_k}+\hat{\chi}_{yy,-2} e^{-2i \theta_k}+
\hat{\chi}_{yy,0} e^{0i \theta_k}\,,\\
\hat{\chi}_{yz}(k_r,\theta_k,k_z)&= i 2^{-1/2} \pi^{3/2} \chi_0 \lambda^5
e^{-\frac{1}{2}\lambda^2(k_r^2+k_z^2)} k_r k_z \left(e^{i \theta_k} -
e^{-i \theta_k}\right)\nonumber\\
&=\hat{\chi}_{yz,1} e^{i \theta_k}+\hat{\chi}_{yz,-1} e^{-i \theta_k}\,.
\end{align}
Starting from the fields $p_r(r,z)$, $p_\theta(r,z)$ and $p_z(r,z)$ which will
numerically be available on a regular grid in $[0, r_{\text{max}}] \times [0,
z_{\text{max}}]$, we first extend all components of $p$ onto $[0, r_{\text{max}}]
\times [-z_{\text{max}}, z_{\text{max}}]$ by reflection symmetry and perform a
Fourier transform in $z$ via FFT to obtain $\tilde{p}_r(r,k_z)$,
$\tilde{p}_\theta(r,k_z)$ and $\tilde{p}_z(r,k_z)$ at regularly spaced Fourier
modes $k_z = 0, \dots, n_z / 2 -1 , -n_z / 2, \dots, -1$. Since $\chi$ decays
exponentially with $k_z^2$, it then suffices to perform the computations described
in the following on a reduced grid only for modes with $|k_z| < k_{z,\text{max}}
\ll n_z / 2$. From these fields, we compute
\begin{align}
\tilde{p}_x(r,\theta,k_z)&=\tilde{p}_r(r,k_z) \cos \theta - \tilde{p}_\theta(r,k_z)
\sin \theta = \frac{1}{2} \left\{( \tilde{p}_r + i \tilde{p}_\theta) e^{i \theta}
+ ( \tilde{p}_r - i \tilde{p}_\theta) e^{-i \theta}\right\}\\
\tilde{p}_y(r,\theta,k_z)&=\tilde{p}_r(r,k_z) \sin \theta + \tilde{p}_\theta(r,k_z)
\cos \theta = \frac{1}{2} \left\{( \tilde{p}_\theta - i \tilde{p}_r) e^{i \theta}
+ ( \tilde{p}_\theta + i \tilde{p}_r) e^{-i \theta}\right\}\\
\tilde{p}_z(r,\theta,k_z) &= \tilde{p}_z(r,k_z) e^{0 i \theta}\,,
\end{align}
or, after numerically evaluating Hankel transforms of $0$th and $1$st order
as detailed below, all appearing Fourier coefficients $\hat{p}_{x,1}(k_r,k_z)$,
$\hat{p}_{x,-1}(k_r, k_z)$, $\hat{p}_{y,1}(k_r,k_z)$, $\hat{p}_{y,-1}(k_r, k_z)$
and $\hat{p}_{z,0}(k_r,k_z)$ where a non-uniform grid in $k_r$ will turn out to
be useful. Then, we calculate the pointwise multiplication of $\hat{\chi}_{yx}$
and $\hat{p}_{x}$ and so on, and arrange the results again in terms of angular
Fourier series
\begin{align}
\hat{\chi}_{yx} \hat{p}_x &= \hat{\chi}_{yx,2} \hat{p}_{x,1} e^{3 i \theta_k} +
\hat{\chi}_{yx,-2} \hat{p}_{x,-1} e^{-3i\theta_k} + \hat{\chi}_{yx,2}
\hat{p}_{x,-1} e^{i\theta_k} + \hat{\chi}_{yx,-2} \hat{p}_{x,1}
e^{-i\theta_k}\,, \label{eq:chip-polar-1}\\
\hat{\chi}_{yy} \hat{p}_y &= \hat{\chi}_{yy,2} \hat{p}_{y,1} e^{3i\theta_k}
+ \hat{\chi}_{yy,-2} \hat{p}_{y,-1} e^{-3i\theta_k} +(\hat{\chi}_{yy,2}
\hat{p}_{y,-1}+\hat{\chi}_{yy,0} \hat{p}_{y,1}) e^{i\theta_k} \nonumber\\
&+(\hat{\chi}_{yy,-2} \hat{p}_{y,1} + \hat{\chi}_{yy,0} \hat{p}_{y,-1})
e^{-i\theta_k\,,} \label{eq:chip-polar-2}\\
\hat{\chi}_{yz} \hat{p}_z &= \hat{\chi}_{yz,1}\hat{p}_{z,0} e^{i \theta_k}
+\hat{\chi}_{yz,-1}\hat{p}_{z,0} e^{-i \theta_k} \,. \label{eq:chip-polar-3}
\end{align}
According to~(\ref{eq:ift-polar}), we then need to compute Hankel transforms
of order $1$ and $3$ on the coefficients, zero-pad all fields to the full $k_z$
grid again and perform an IFFT in $z$, plug in $\theta = 0$ in the end and add
all contributions in order to obtain the result~$(\chi * p)_\theta$. The efficiency
of the proposed method hence depends on two ingredients: The fact that only a small
number of angular modes appears for our choice of $\chi$, which limits the number
of different order Hankel transforms that have to be computed, plus an efficient
implementation of the Hankel transform evaluations themselves.

Following~\cite{sneddon:1995,fisk-johnson:1987,melchert-wollweber-roth:2018}, we
use the following series expansion with fast convergence to compute the
\textit{inverse} Hankel transform from Fourier space to real space:
\begin{align}
\tilde{f}_l(r,k_z) \approx \frac{i^l}{2\pi} \frac{2}{r_\text{max}^2}
\sum_{m = 1}^{n_{\text{Bz}}} \frac{J_n\left(j_{l,m} \, r/r_\text{max}\right)}
{(J_{l}'(j_{l,m}))^2} \hat{f}_l(j_{l,m}/r_\text{max},k_z)\,, \label{eq:hankelbkwd}
\end{align}
where $j_{l,m}$ denotes the $m$-th zero (in ascending order) of the Bessel
function $J_l$. For this, we need to evaluate the \textit{forward} Hankel
transforms of the components of $p$ at $n_{\text{Bz}} \ll n_r$ points $j_{l,m}/
r_{\text{max}}$ for all angular modes $l$ that appear in the
products~(\ref{eq:chip-polar-1}) to~(\ref{eq:chip-polar-3}). We calculate these
forward transforms of the components of $p$ with a straightforward quadrature
of~(\ref{eq:def-hankel}), specifically with the trapezoidal rule as
in~\cite{melchert-wollweber-roth:2018}.

\subsection{Numerical implementation and parameters}

For the axisymmetric instanton equations, a simplified version of the gradient
descent algorithm~\ref{algo:gradient-descent} turned out to be sufficient.
We only specify Lagrange multipliers ${\cal F}$ without penalty, perform
updates~(\ref{eq:cs-update}) with fixed step sizes that are reduced by a
factor $\tilde{\beta}$ whenever an oscillation of $O[u[p](\cdot, 0)]$ between
iterations is detected, and stop iterating once the relative change of
$O[u[p](\cdot, 0)]$ between iterations drops below a threshold $\delta$
according to
\begin{align}
\left| \frac{O\left[u\left[p^{(k)}\right](\cdot, 0)\right]
- O\left[u\left[p^{(k-1)}\right](\cdot, 0)\right]}{\sigma^{(k-1)}
O\left[u\left[p^{(k-1)}\right](\cdot, 0)\right]}  \right| < \delta\,.
\label{eq:stopping-cs}
\end{align}
The numerical parameters for the axisymmetric solver that have been used to obtain
the results of the main paper are summarised in table~\ref{tab:cylin-params}.
Regarding the concrete implementation, the code that has been written and used
for this task consists of Fortran 90 subroutines for the Leapfrog steps and
the multigrid solver that are called in Python via F2PY~\cite{peterson:2009}.

\begin{table}
\caption{Parameters of the axisymmetric (2+1)-dimensional solver
of the instanton equations~(\ref{eq:inst-cylin-supmat}) using a
forward-backward iteration at fixed values of ${\cal F}$.}
\label{tab:cylin-params}
\centering
\begin{tabular}{L{3cm}L{5cm}R{3cm}}
\toprule
\textbf{Parameter} & \textbf{Explanation} & \textbf{Typical value}\\
\midrule 
$T$ & Time interval $[-T,0]$ & $1$\\
$r_{\text{max}}$ & Radial domain $[0, r_{\text{max}}]$ & $\pi$\\
$z_{\text{max}}$ & $z$-domain $[-z_{\text{max}}, z_{\text{max}}]$ & $\pi$\\
$u_0$   & Fixed initial condition for $u$ & $0$\\
$\chi_0$ & Default forcing strength in~(\ref{eq:forcing-matrix}) & $1$\\
$\lambda$ & Forcing correlation length in~(\ref{eq:forcing-matrix}) & $1$\\
$n_r$, $n_z$ & Resolution of $[0, r_{\text{max}}] \times [0, z_{\text{max}}]$ & $256$\\
$n_t$ & Temporal resolution & $1024$\\
$t_\alpha$ & Stretching factor for time in~(\ref{eq:time-parameterization})
& $3$\\
$n_{z,\text{red}}$ & Reduced number of $z$ modes & $10$\\
$n_{\text{Bz}}$ & \# of Bessel zeros for convolutions & $10$\\
$R$ & Width of the final conditions for $p_{\text{I}}$ & $\Delta r$\\
$\tilde{\beta}$ & Step size reduction factor & $0.8$\\
$\delta$ & Threshold for stopping in~(\ref{eq:stopping-cs}) & $5 \cdot 10^{-3}$\\
\bottomrule
\end{tabular}
\end{table}


\bibliography{bib}

\end{document}